\documentclass[aps,prd,preprint,superscriptaddress,nofootinbib]{revtex4-2}
\usepackage{graphicx} 

\usepackage{amsmath,amssymb}
\usepackage{hyperref}
\usepackage{booktabs}
\usepackage{xcolor}
\usepackage{subcaption}
\usepackage{tikz}
\usepackage{float}
\usepackage[compat=1.1.0]{tikz-feynman}
\usepackage{hepunits}   
\usepackage{cleveref}
\newcommand{\ma}{m_a}
\newcommand{\gagg}{g_{a\gamma\gamma}}

\newcommand{\dr}{\Delta R}
\newcommand{\sqrts}{\sqrt{s}}

\newcommand{\fbinv}{$\,\text{fb}^{-1}$}

\newcommand{\kl}{$K_L $ }

\newcommand{\GeV}{\,\text{GeV}}   
\newcommand{\TeV}{\,\text{TeV}}   
\newcommand{\MeV}{\,\text{MeV}}   
\newcommand{\ee}{e^+e^-}

\hypersetup{
    colorlinks=true,
    linkcolor=blue,
    citecolor=blue,
    urlcolor=blue
}
\usepackage[normalem]{ulem}

\begin{document}

\title{\textbf{Probing light axion-like particles in vector boson fusion at CMS with data parking and scouting}}
\author{Sena Durgut}
\affiliation{Department of Physics, Carnegie Mellon University, Pittsburgh, Pennsylvania 15213, USA}
\author{Mariel Peczak}
\affiliation{Department of Physics and Astronomy, University of Pennsylvania, Philadelphia, Pennsylvania 19104, USA}
\author{Gonzalo Alonso-Álvarez}
\affiliation{Instituto Galego de Física de Altas Enerxías, Universidade de Santiago de Compostela, 15782 Santiago de Compostela, Galicia, Spain}
\author{Chiara Amendola}
\affiliation{Department of Physics, Carnegie Mellon University, Pittsburgh, Pennsylvania 15213, USA}
\author{Matteo Cremonesi}
\affiliation{Department of Physics, Carnegie Mellon University, Pittsburgh, Pennsylvania 15213, USA}
\author{Joerg Jaeckel} 
\affiliation{Institut für Theoretische Physik, Universität Heidelberg, Philosophenweg 16, 69120 Heidelberg, Germany}
\author{Matteo Marchegiani}
\affiliation{Department of Physics, Carnegie Mellon University, Pittsburgh, Pennsylvania 15213, USA}
\date{\today}

\begin{abstract}
Axion-like particles (ALPs) with masses between 10 MeV and 10 GeV and moderately small couplings sit in an experimental blind spot. They decay too fast for intensity-frontier experiments and are too feebly coupled and too light for conventional collider searches. The LHC produces them at substantial rates through vector boson fusion (VBF), the dominant production mode for electromagnetically coupled ALPs, but two obstacles keep this signal out of reach. The events are soft, so trigger thresholds discard most of them, and the Lorentz boost merges the two decay photons into a single calorimeter deposit. We show that the Compact Muon Solenoid (CMS) experiment can overcome both by combining its data-acquisition strategies with the tracker-based reconstruction of merged photon pairs. Data parking on the VBF jet topology in Run 3 (312 fb$^{-1}$) and trigger-level data scouting at the High-Luminosity LHC (HL-LHC, 3 ab$^{-1}$) remove the need for a threshold on the photons. Photons that convert to electron-positron pairs in the silicon tracker resolve the merged pair, since the tracker measures their direction far more finely than the calorimeter granularity. Reconstructing the ALP decay vertex from the conversion tracks removes the prompt backgrounds but not the long-lived $K_L\to\gamma\gamma$. In Run 3, where the full event is recorded, the merged diphoton invariant mass confines the $K_L$ to a narrow window around its own mass. In the scouting stream, where only trigger-level information survives, the diphoton $p_T$, the conversion tracks, and the hadronic activity in the event take over this role. We project that the parked Run 3 data, already recorded, reach a region of the $(m_a, g_{a\gamma\gamma})$ plane that no measurement has probed. Level-1 trigger data scouting at the HL-LHC extends that reach by roughly an order of magnitude in both coupling and mass.
\end{abstract}
\maketitle

\section{Introduction}
\label{sec:intro}


The exploration of new physics beyond the Standard Model (SM) is pushed forward both at the high-energy and at the high-precision and intensity frontiers. The former traditionally searches for heavy particles with SM-strength couplings, while the latter is sensitive mostly to light and very feebly interacting particles. Just in between, at masses between roughly 10~MeV and 10~GeV and couplings of about $10^{-3}$--$10^{-5}$~GeV$^{-1}$, there is a gap in sensitivity (cf., e.g.~\cite{Jaeckel:2015jla,Beacham:2019nyx,Agrawal:2021dbo,Antel:2023hkf}).

Experiments operating at those intermediate energies face important challenges. Fixed-target experiments such as the upcoming SHiP experiment~\cite{SHiP:2015vad,Alekhin:2015byh} are limited by the short lifetimes of moderately coupled
states, whose decay products are absorbed in shielding before they can be
detected, and by their limited reach above the GeV scale. Low-energy,
high-luminosity $\ee$ machines such as Belle~II are promising but cover only sufficiently large couplings, leaving the lower coupling region unexplored \cite{Kou:2018nap}. In contrast, high-energy hadron colliders, such as the Large Hadron Collider (LHC), offer both ample energy and the integrated luminosity needed to produce such states. The first difficulty in these experiments is the identification of the signal against the background. The second obstacle is the trigger. The LHC produces data at rates far exceeding what can be stored, so  events are selected online, typically setting thresholds on energy, angular position, or topological features. Low-mass signals with soft, low-multiplicity final states are discarded by the majority of trigger protocols.
 
Axion-like particles (ALPs)~\cite{Masso:1995tw,Masso:1997ru} (cf., e.g.~\cite{Jaeckel:2010ni,Beacham:2019nyx,Agrawal:2021dbo,Adams:2022pbo,Antel:2023hkf,Albertus:2026fbe,Arza:2026rsl} for reviews) provide a concrete example of a hard-to-detect feebly-interacting-particle. They are a natural generalization of the standard quantum chromodynamics (QCD) axion~\cite{Peccei:1977hh,Weinberg:1977ma,Wilczek:1977pj}, which is deeply motivated by the strong charge-parity problem and remains a leading cold dark matter candidate~\cite{Preskill:1982cy,Abbott:1982af,Dine:1982ah}. Unlike the standard QCD axion, ALP masses and decay constants are independent model parameters. ALPs arise generically as pseudo-Nambu-Goldstone bosons of spontaneously broken global U(1) symmetries, cf., e.g.~\cite{Masso:1995tw,Jaeckel:2010ni,Beacham:2019nyx,Agrawal:2021dbo,pdg}. 

From a theoretical perspective, there is a strong motivation to push the sensitivity to ALPs in the GeV region. Even though the canonical QCD-axion photon-coupling band for this mass range lies well above current constraints, models exist in which the QCD axion is much heavier than the canonical expectation due to the contribution of small-scale instantons~\cite{Agrawal:2017ksf,Agrawal:2017evu,Gaillard:2018xgk,Buen-Abad:2019uoc}.
These can populate the region of parameter space under study in this work. 
In a general ALP scenario, the ALP couplings are suppressed by the symmetry breaking scale $f_a$, which often also acts as the cutoff of the effective ALP description~\cite{Brivio:2017ije,Bauer:2017ris,Alonso-Alvarez:2018irt,Bauer:2020jbp} (see~\cite{Alonso-Alvarez:2021ett} for a quantitative discussion).
ALP-photon couplings typically scale as $g_{a\gamma\gamma}=\alpha /(2\pi ) \cdot C_\gamma / f_a$, where $\alpha$ is the fine-structure constant and $C_\gamma$ is a UV-model parameter that is expected to be $\mathcal{O}(1)$ in minimal constructions.
Thus, testing a given ALP-photon coupling indirectly probes a new physics scale
\begin{align}
    f_a \simeq \mathrm{TeV}\,\left( \frac{10^{-6}\,\mathrm{GeV}^{-1}}{g_{a\gamma\gamma}} \right) \left(\frac{C_\gamma}{1}\right),
\end{align}
showcasing the need to reach smaller couplings to push towards the TeV frontier and beyond.
Furthermore, in order to have a consistent ALP description as a weakly-coupled pseudo-Nambu-Goldstone boson, the ALP mass should lie below this high energy scale, which translates into a requirement 
\begin{equation}
    m_a \ll 10\,\mathrm{GeV}\times \left(\frac{10^{-4}\,\mathrm{GeV}^{-1}}{g_{a\gamma\gamma}}\right),
\end{equation} 
for $C_\gamma=1$\footnote{The expectation of $C_\gamma \sim 1$ can be broken in non-minimal models with enhanced or suppressed anomaly coefficients~\cite{Craig:2018kne}, like clockwork~\cite{Giudice:2016yja,Coy:2017yex} or alignment~\cite{Agrawal:2017cmd} mechanisms or composite UV completions~\cite{Gripaios:2009pe,Redi:2016esr}.}.
Given the sensitivity of current ALP searches at high-energy colliders, it is clear that satisfying this theoretical consistency condition motivates developing strategies that can target the challenging sub-10 GeV regime.

From an experimental perspective, an electromagnetically coupled ALP would manifest as a diphoton resonance~\cite{Jaeckel:2012yz}, in close analogy with the Higgs boson~\cite{higgs}. However, searching for such a signal in the 10~MeV--10~GeV mass range presents two important challenges. First, SM mesons produce copious backgrounds, either through their own diphoton decays or through decay products that can be misreconstructed as photons. Second, ALPs produced at the LHC in this mass range are typically highly boosted, and the two decay photons are correspondingly
collimated. At sufficiently low masses, their angular separation falls below the granularity of the electromagnetic calorimeter (ECAL), causing the two photons to deposit their energy in the same cell. The resulting merged electromagnetic shower can then mimic that of a single photon, making conventional diphoton searches ineffective in this regime.

A strategy to overcome these challenges was proposed in
Ref.~\cite{tracking-alps}, building on the earlier
observation~\cite{Dasgupta_2016} that collimated diphotons can be
distinguished by their enhanced photon-conversion probability. The key idea is
to exploit the fraction of decay photons that convert to $\ee$ pairs within the
tracker material. Because the tracker resolves the resulting conversion tracks
with an angular
precision that is two orders of magnitude finer than the ECAL granularity, the two collimated photons can be separated and the 
displaced ALP decay vertex can
be reconstructed directly. Applied to ALPs produced via vector boson fusion (VBF) in the ATLAS detector, this approach was shown to access substantial
untested regions of the $(\ma,\,\gagg)$ plane. That study, however, was restricted to events passing a $140\GeV$ single-photon trigger,
which discards around $75\%$ of the signal. 

In the present note, we explore strategies to address these challenges by combining theoretical ideas from~\cite{tracking-alps} with data-acquisition strategies developed by the  Compact Muon Solenoid (CMS) collaboration, namely data parking and data scouting \cite{scouting-parking}. With this, we project the sensitivity of the CMS experiment to light VBF-produced ALPs
decaying to collimated photon pairs at $\sqrts = 13.6$~TeV and 14~TeV, following a staged strategy. For Run~3, we exploit data parking, which defers the reconstruction of events stored
in raw format. This allows events to be selected on the VBF jet topology alone, with
no threshold on the ALP decay products. The full event content is
preserved, so the invariant mass of the merged diphoton system can be reconstructed using machine learning (ML) from the ECAL
shower and the tracks. At the High-Luminosity LHC (HL-LHC), we use data scouting, which records only a compact summary of
each event as reconstructed in the trigger. Events are retained at the full
bunch-crossing rate without thresholds, at the cost of the invariant-mass
reconstruction. We follow the tracking-conversion strategy of Ref.~\cite{tracking-alps} to reconstruct the displaced ALP vertex and reject prompt backgrounds in both cases. The relaxed thresholds also admit a long-lived background, $K_L\to\gamma\gamma$, that survives the displacement requirement, and we discuss how the invariant mass in Run~3 and the diphoton $p_T$ in the scouting stream can be used to suppress it. We find that this approach can unlock a portion of the parameter space that has not been explored before. The breadth of trigger and
acquisition strategies available to CMS is what drives the projected reach.

\section{Benchmark ALP model and production}
\label{sec:model}

We adopt a standard benchmark model (cf., e.g.~\cite{Agrawal:2021dbo}) of an ALP coupled to the electroweak gauge
bosons. In the following, we summarize the resulting decay and lifetime, and describe the VBF
production and its simulation. 

\subsection{Effective Lagrangian and decay}
\label{sec:lagrangian} 
An ALP coupling to photons alone is described by
\begin{equation}
\mathcal{L} \supset \tfrac{1}{2}(\partial_\mu a)^2 - \tfrac{1}{2}\ma^2 a^2
- \tfrac{1}{4}\,\gagg\, a\, F_{\mu\nu}\tilde{F}^{\mu\nu},
\label{eq:lag_gamma}
\end{equation}
with $\tilde{F}^{\mu\nu}=\tfrac{1}{2}\epsilon^{\mu\nu\rho\sigma}F_{\rho\sigma}$.
Since electroweak gauge bosons also participate in the VBF process, their couplings must be specified too. A simple choice is to start
from a coupling to the hypercharge field strength $B_{\mu\nu}$,
\begin{equation}
\mathcal{L} \supset -\tfrac{1}{4}\frac{\gagg}{c_W^2}\, a\, B_{\mu\nu}\tilde{B}^{\mu\nu}
= -\tfrac{1}{4}\gagg\, a\,F\tilde{F}
  -\tfrac{1}{4}\frac{s_W^2}{c_W^2}\gagg\, a\,Z\tilde{Z}
  +\tfrac{1}{4}\frac{2 s_W}{c_W}\gagg\, a\,F\tilde{Z},
\label{eq:lag_hyper}
\end{equation}
which after electroweak symmetry breaking generates the $\gamma\gamma$,
$\gamma Z$ and $ZZ$ couplings simultaneously, normalized to reproduce the photon
coupling of Eq.~\eqref{eq:lag_gamma}~\cite{tracking-alps}. Here $s_W,c_W$ denote the sine and cosine of the Weinberg angle. Coupling to
$B_{\mu\nu}$ alone is a choice, couplings to $W^a_{\mu\nu}$, or to a combination,
reproduce $\gagg$ under a suitable normalization but leave the $\gamma Z$ and $ZZ$
couplings unfixed. We adopt the hypercharge case as a benchmark, following
Ref.~\cite{tracking-alps}.

In the mass range of interest, ALPs decay predominantly to two
photons, with width
\begin{equation}
\Gamma(a\to\gamma\gamma) = \frac{\gagg^2\,\ma^3}{64\pi}.
\label{eq:width}
\end{equation}
The boosted, lab-frame decay length is
$\ell_{\rm decay} = (p_a/\ma)\,c\tau$ with $c\tau = 1/\Gamma$, which numerically scales as
\begin{equation}
\ell_{\rm decay} \approx 40~\mathrm{cm}
\left(\frac{p_a}{150\GeV}\right)
\left(\frac{0.2\GeV}{\ma}\right)^{4}
\left(\frac{0.1\TeV^{-1}}{\gagg}\right)^{2}.
\label{eq:declength}
\end{equation}
For the masses and couplings that we target, $\ell_{\rm decay}$ ranges from millimeters to meters and is thus macroscopic on
collider scales. The displacement of the decay vertex
from the primary vertex is one of the main discriminants that we exploit to distinguish ALPs from promptly decaying mesons (Sec.~\ref{sec:bkg}).
\subsection{Production via vector boson fusion}
\label{sec:production}
\begin{figure}
\centering
\begin{tikzpicture}
  \begin{feynman}
    \vertex (q1i) at (-3, 2) {\(q\)};
    \vertex (q2i) at (-3,-2) {\(q\)};
    \vertex (v1)  at (-1, 1.5);
    \vertex (v2)  at (-1,-1.5);
    \vertex (q1o) at ( 1, 2.5) {\(q\)};
    \vertex (q2o) at ( 1,-2.5) {\(q\)};
    \vertex (vc)  at ( 1, 0);        
    \vertex (va)  at ( 3, 0);        
    \vertex (g1)  at ( 5, 1) {\(\gamma\)};
    \vertex (g2)  at ( 5,-1) {\(\gamma\)};

    \diagram* {
      (q1i) -- [fermion, thick] (v1) -- [fermion, thick] (q1o),
      (q2i) -- [fermion, thick] (v2) -- [fermion, thick] (q2o),
      (v1) -- [boson, edge label=\(\gamma/Z\)] (vc),
      (v2) -- [boson, edge label'=\(\gamma/Z\)] (vc),
      (vc) -- [scalar, edge label=\(a\)] (va),
      (va) -- [boson] (g1),
      (va) -- [boson] (g2),
    };
  \end{feynman}
\end{tikzpicture}
\caption{VBF production of an electromagnetically coupled ALP
in $pp$ collisions, followed by $a\to\gamma\gamma$.}
\label{fig:vbf}
\end{figure}
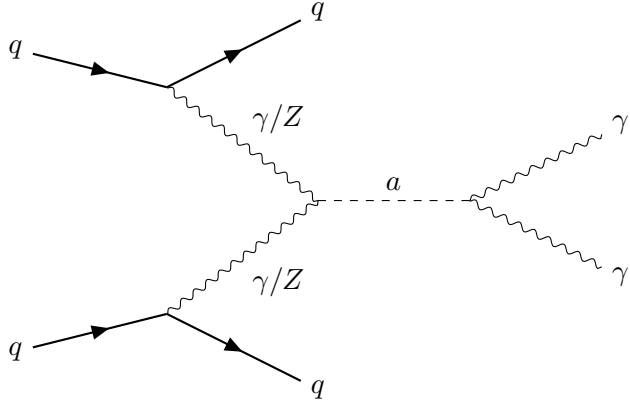

The leading production mode for electromagnetically coupled ALPs at the LHC is VBF, $pp \to a\,jj$, in which two quarks each radiate an
electroweak boson ($\gamma/Z$) that fuse to produce the ALP, leaving two forward jets. The diagram corresponding to this process is shown in Fig.~\ref{fig:vbf}. At high partonic energies, the $\gamma\gamma$,
$\gamma Z$ and $ZZ$ channels contribute at comparable levels.

To leading order, the inclusive VBF cross section scales quadratically with the
coupling and is, to a good approximation, independent of $\ma$ across the
$10\MeV$--$10\GeV$ window \cite{tracking-alps},
\begin{equation}
\sigma(pp\to a\,jj) \approx 180~\mathrm{pb}
\left(\frac{\gagg}{10^{-2}\GeV^{-1}}\right)^{2}.
\label{eq:xsec}
\end{equation}

\subsection{Signal simulation}
\label{sec:simulation}
Signal samples are generated at leading order with
\textsc{MadGraph5\_aMC@NLO}~v3.5.13~\cite{madgraph} using a custom implementation of the ALP effective
model of Ref.~\cite{tracking-alps}, for the VBF process
$pp \to a\,jj$ ($a\to\gamma\gamma$) at $\sqrts = 13.6\TeV$. We generate mass points spanning
$10\MeV$--$10\GeV$. No generation cuts were applied at this stage and all selections are imposed at
the analysis level.

\begin{figure}
    \centering
    \begin{subfigure}[b]{0.49\linewidth}
        \includegraphics[width=\linewidth]{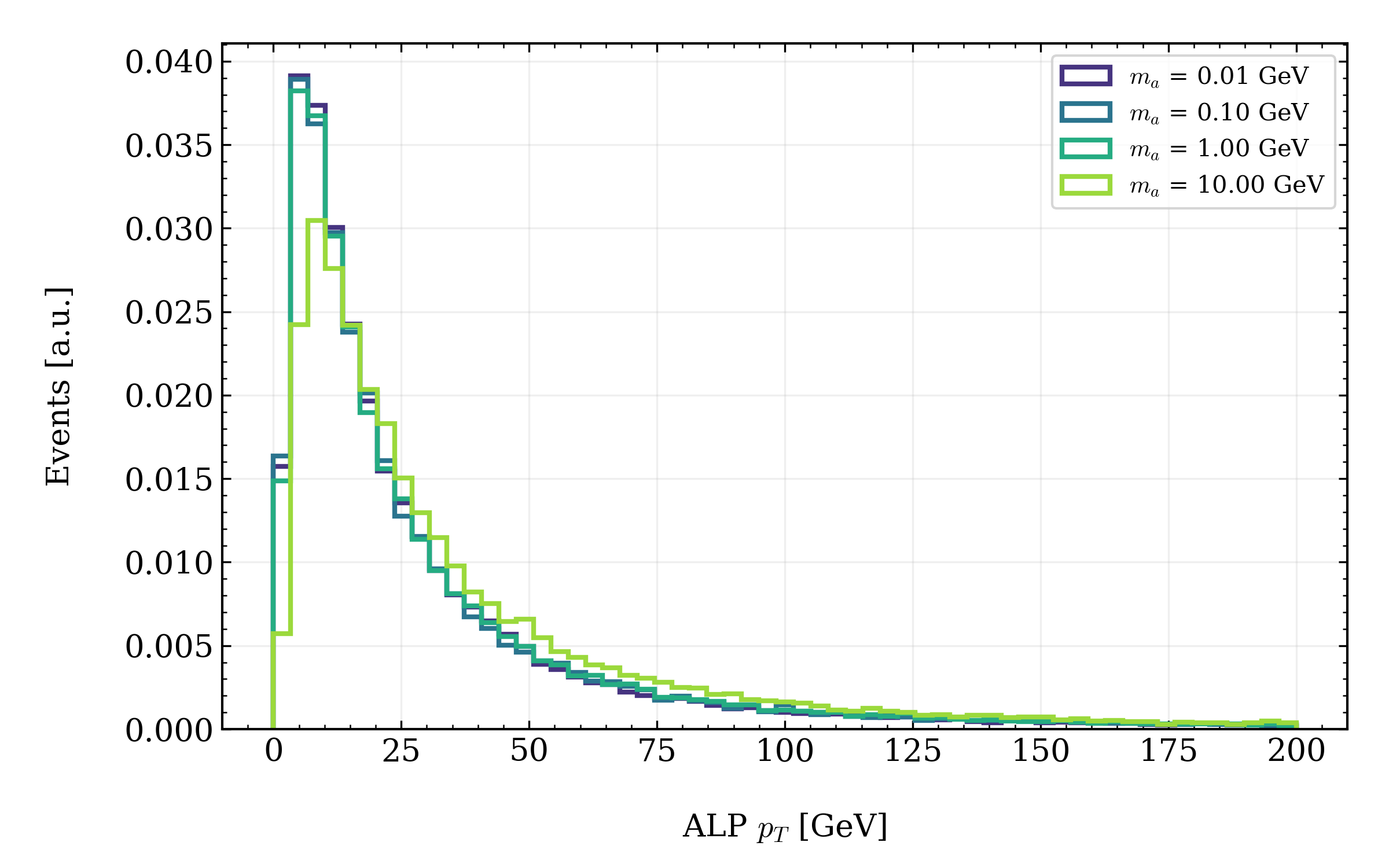}
        \caption{}
        \label{fig:alp_pt}
    \end{subfigure}%
    \hfill
    \begin{subfigure}[b]{0.49\linewidth}
        \includegraphics[width=\linewidth]{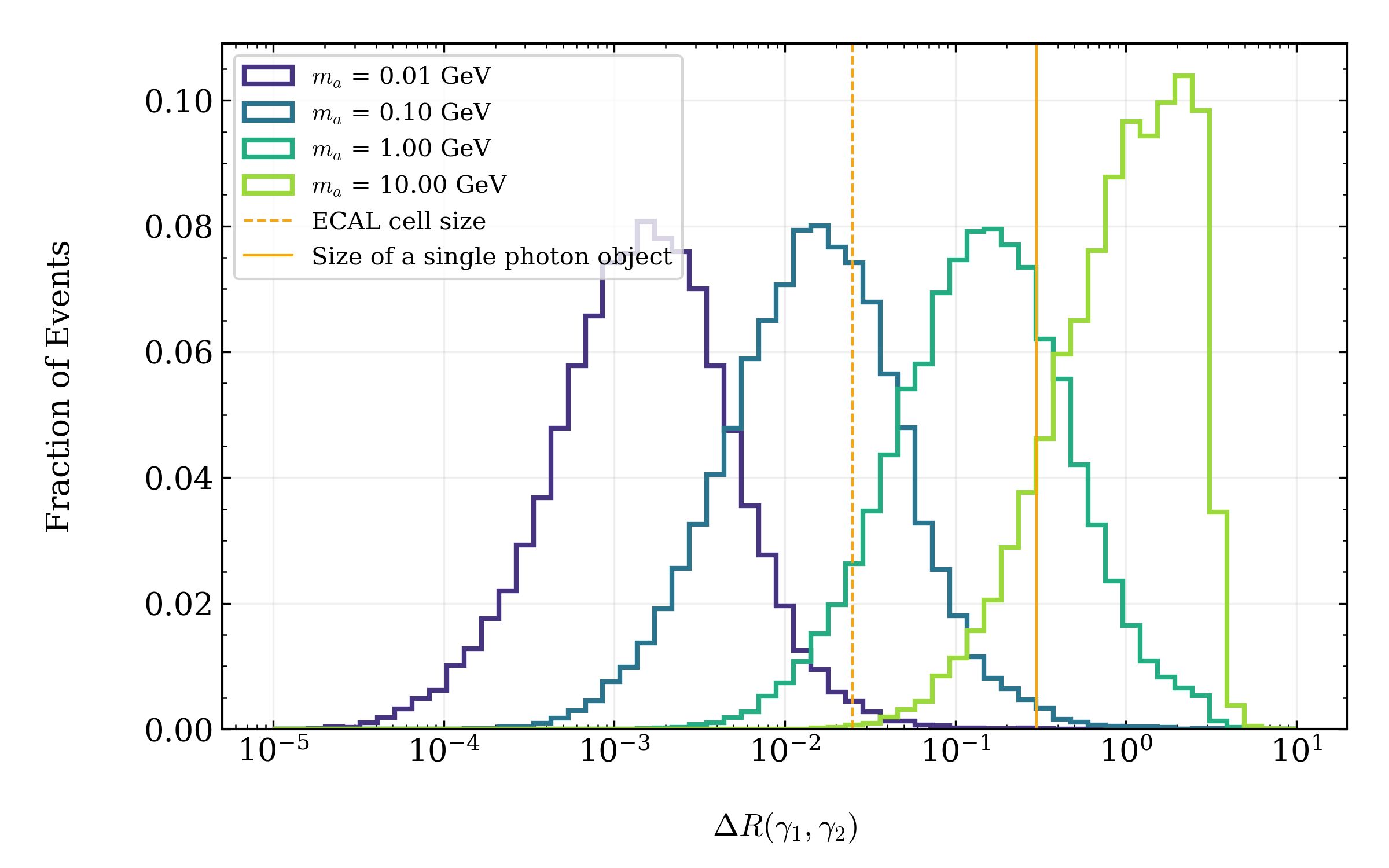}
        \caption{}
        \label{fig:dr_dist}
    \end{subfigure}
    
    \vspace{0.5cm} 
    
    \begin{subfigure}[b]{0.49\linewidth}
        \includegraphics[width=\linewidth]{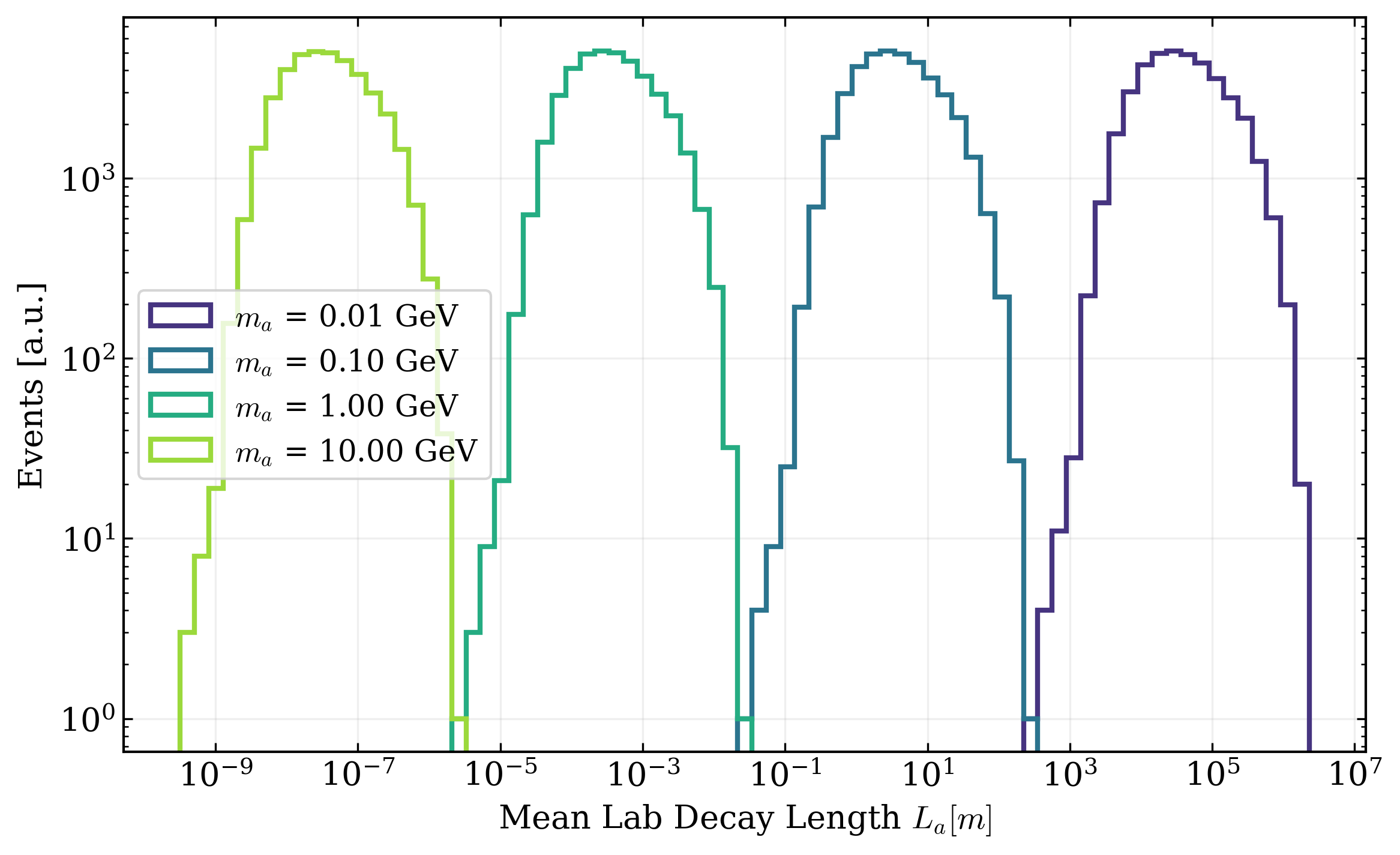}
        \caption{}
        \label{fig:alp_decay_distr}
    \end{subfigure}
    \caption{Signal kinematic distributions for ALPs produced in  $pp$ collisions at
    $\sqrt{s}=13.6\text{ TeV}$ and fixed $g_{a\gamma\gamma} = 0.1\text{ TeV}^{-1}$, for several ALP masses:
    (a) transverse momentum $p_T$, (b) diphoton opening angle
    $\Delta R(\gamma_1,\gamma_2)$, and (c) lab-frame mean decay length.}
    \label{fig:signal_kinematics}
\end{figure}

In Fig.~\ref{fig:signal_kinematics} we summarize the signal kinematics for four
representative masses, $\ma \in \{0.01, 0.1, 1, 10\}\GeV$, at fixed
$\gagg = 0.1\TeV^{-1}$. The ALP transverse momentum
(Fig.~\ref{fig:signal_kinematics}a) is soft and nearly mass independent. It peaks
around $p_T \sim 5\text{--}10\GeV$ and falls steeply, with only a small tail above
$100\GeV$. Most of the signal therefore sits well below a typical single-photon
trigger threshold, around $140\GeV$ at ATLAS and $200\GeV$ at CMS. Recovering it is what motivates the low-threshold acquisition
strategies described in Sec.~\ref{sec:strategy}.

Sub-GeV ALPs carry a large Lorentz boost, which collimates their decay products. In
the small-angle limit the opening angle between the two photons scales as \cite{tracking-alps}
\begin{equation}
\Delta R \simeq \frac{2\ma}{p_a},
\label{eq:dR}
\end{equation}
and the simulated distribution (Fig.~\ref{fig:signal_kinematics}b) follows this trend.

The photons become more collimated as the mass drops, and the peak moves from
$\dr \sim 2$ at $\ma = 10\GeV$ down to $\dr \sim 10^{-3}$ at $\ma = 10\MeV$. The
conversion tracks serve two purposes in our strategy. They enter the reconstruction of
the displaced ALP decay vertex, which rejects prompt backgrounds and is applied across
the full mass range irrespective of the photon separation. They also recover the
two-photon nature of the signal once the calorimeter alone can no longer do so. Over
most of the mass range the separation falls below the size of a reconstructed photon
object at CMS ($\dr = 0.3$~\cite{Sirunyan_2021}), so the pair is reconstructed as a
single photon. For $\ma \lesssim 0.1\GeV$ it drops further, below the ECAL cell size
($\dr = 0.025$) and the
conversion tracks become the only handle on the pair.

The mean lab decay length (Fig.~\ref{fig:signal_kinematics}c) shows the steep
$\ell_{\rm decay}\propto p_a/\ma^{4}$ dependence of Eq.~\eqref{eq:declength}. For the
lightest masses the mean length lies far beyond the detector, so the signal that
decays inside the tracker comes from the short tail of the exponential. That displaced
vertex is the primary discriminant against prompt backgrounds (Sec.~\ref{sec:bkg}).

\section{General strategy outline}
\label{sec:strategy}
In this section we outline the search strategies that we propose for Run~3 and Phase-2. We
first describe the CMS detector, the photon-conversion mechanism, and the
displaced-vertex reconstruction that our approach relies on. We then explain why
alternative triggering strategies are essential to make the $10\MeV$--$10\GeV$ ALP
mass range accessible, and finally present the Run~3 and Phase-2 strategies.

\subsection{The CMS detector}
\label{sec:detector}
The CMS apparatus is built around a $3.8$~T superconducting solenoid. The
innermost system is an all-silicon tracker (pixel and strip detectors)
providing transverse impact-parameter resolution of
$\mathcal{O}(10\text{--}15~\mu\mathrm{m})$ for high-$p_T$ tracks over
$|\eta|\lesssim2.5$. The Phase-2 upgrade replaces it with a lighter
tracker~\cite{CERN-LHCC-2017-009}. It consists of an Inner Tracker (IT) of silicon
pixel modules, which extends the geometrical acceptance to $|\eta| = 4$, and an Outer
Tracker (OT) of silicon strip modules. The OT uses two module types. Pixel-strip (PS)
modules populate the inner layers, and strip-strip (2S) modules, made of two closely
spaced strip sensors, populate the outer layers. Surrounding the
tracker, the lead-tungstate ECAL provides a barrel ($|\eta|<1.48$) and two endcaps
(to $|\eta|<3$). For the HL-LHC the endcap calorimeters are replaced by the
high-granularity calorimeter (HGCAL), covering $1.4<|\eta|<3$ with fine spatial
segmentation and sub-nanosecond time precision~\cite{phase2_paper}. The HL-LHC will deliver about $3\,\mathrm{ab}^{-1}$, roughly ten times the Run~3
dataset, at the cost of higher pileup and radiation that drive the detector upgrades
described above. 
\subsection{Photon conversions}
\label{sec:conversions}
Standard photon reconstruction and identification at CMS
relies on the ECAL, whose barrel is segmented into crystals of
$\delta\eta\times\delta\phi = 0.0174\times0.0174$, or $\Delta R \simeq 0.025$. The information is fed to the 
Level-1 trigger (L1T) with coarser granularity, set by arrays of
$5\times5$ crystals, $\Delta\eta\times\Delta\phi \simeq 0.087\times0.087$, or
$\Delta R \simeq 0.12$. Comparing these scales with the photon-pair separation of
Fig.~\ref{fig:signal_kinematics}b shows that, for $\ma \lesssim 1\GeV$, two
photons are too close to be resolved for trigger selection and they mimic a single photon.

\begin{figure}
    \centering
    \begin{subfigure}{0.49\linewidth}
        \includegraphics[width=\linewidth]{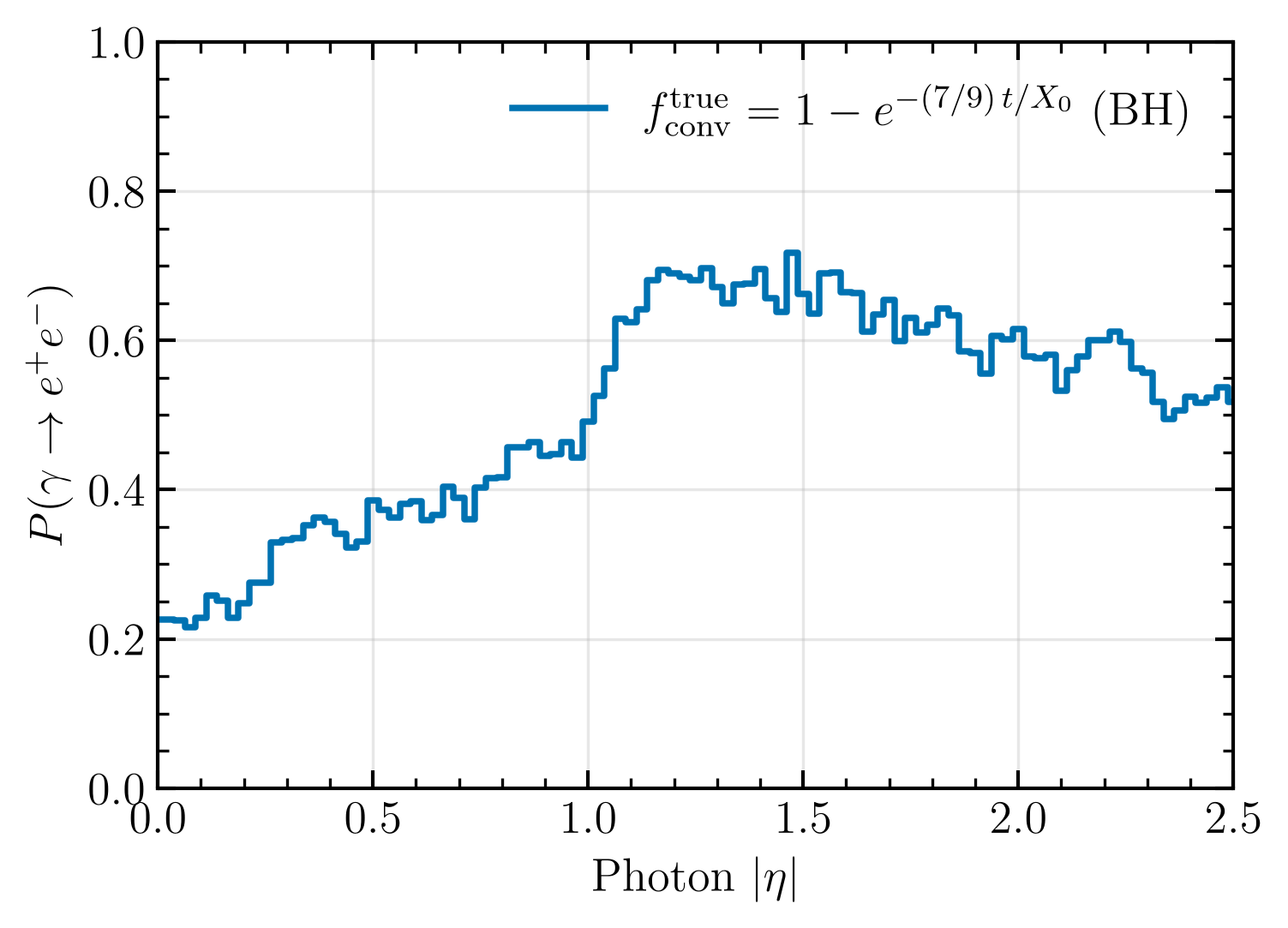}
        \caption{Phase-1}
        \label{fig:phase1}
    \end{subfigure}%
    \hfill
    \begin{subfigure}{0.49\linewidth}
        \includegraphics[width=\linewidth]{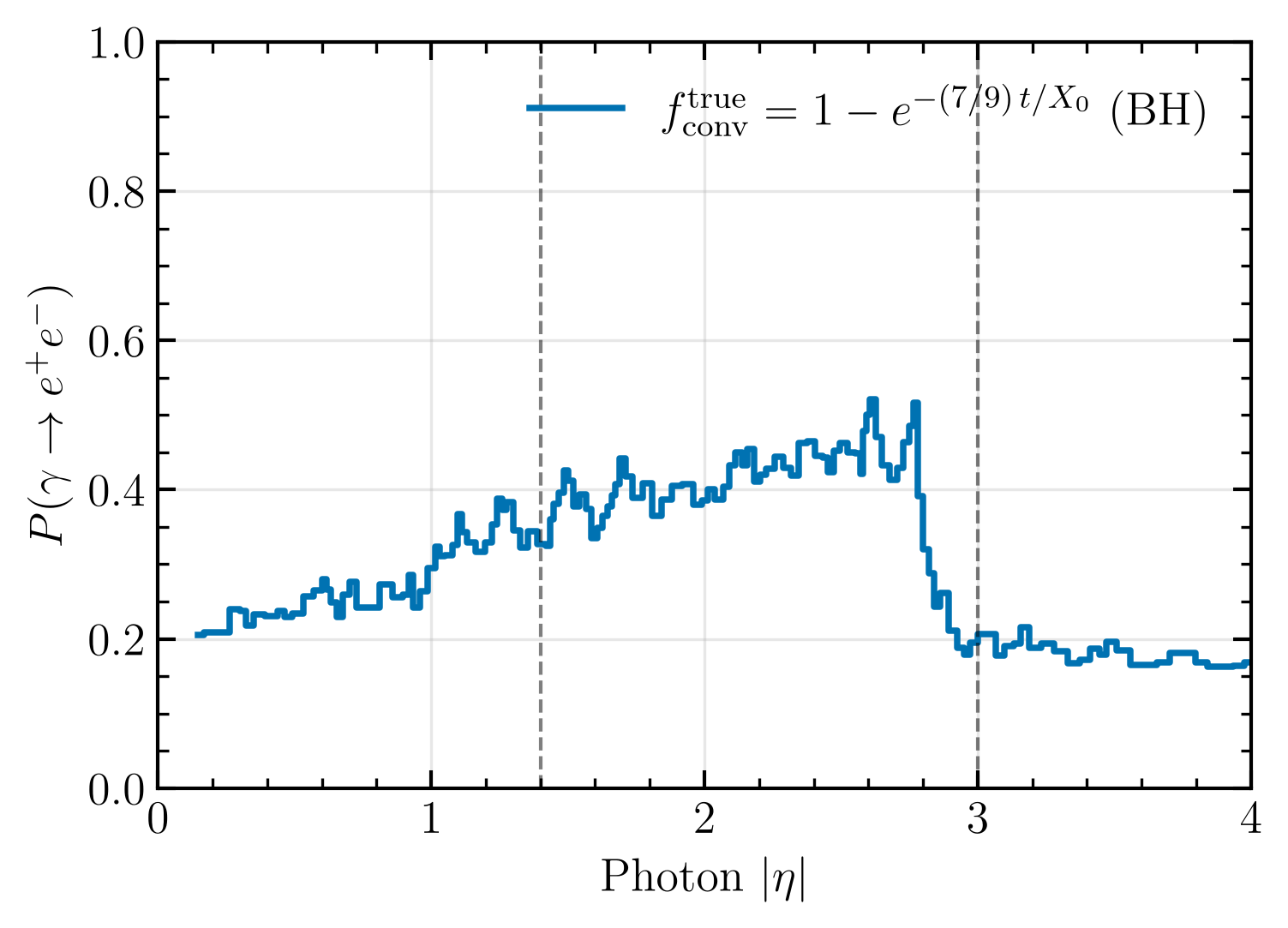}
        \caption{Phase-2}
        \label{fig:phase2}
    \end{subfigure}
    \caption{Derived photon conversion probability, $P_{\text{conv}}$, as a function of pseudorapidity $\eta$ across the CMS tracker acceptance. The probabilities are calculated using Eq.~\eqref{eq:conv_prob_t} based on the tracker material budget profiles taken from \cite{CERN-LHCC-2017-009} for the (a) Phase-1 and (b) Phase-2 detectors. The dashed lines in the Phase-2 distribution indicate the High Granularity Calorimeter (HGCAL) region.}
    \label{fig:conv-prob-combined} 
\end{figure}

Photon conversions provide a way around this. A photon traversing the tracker material
can convert into an $e^+e^-$ pair through the Bethe–Heitler process. For high-energy photons, the conversion proceeds
with negligible energy or momentum transfer, so the $e^+e^-$ pair carries the momentum of the
parent photon. The photon direction can therefore be recovered from the tracks of the $e^+e^-$ pair. A sizable fraction of photons convert before leaving the CMS tracker, up
to 80\% in the existing setup (Fig.~\ref{fig:phase1}) and 50\% in Phase-2 (Fig.~\ref{fig:phase2})  depending on pseudorapidity, so it is likely that both photons from a
light-ALP decay convert. The tracker resolves
the conversion tracks with an angular resolution more than two orders of magnitude finer
than the calorimeter granularity, so a collimated pair that merges into a single calorimeter
object can still be separated from the tracks alone, and the ALP decay vertex
reconstructed from them.

The conversion probability is set by the tracker material budget. In the high-energy
limit ($E_\gamma \gg m_e$) the photon mean free path is
$\lambda_{\text{pair}} = (9/7)\,X_0$~\cite{pdg}, so a photon traversing a depth of
$t \equiv x/X_0$ radiation lengths converts with probability
\begin{equation}
  P_{\text{conv}}(t) = 1 - e^{-7t/9} .
  \label{eq:conv_prob_t}
\end{equation}
Based on the published CMS material budget $t(\eta)$~\cite{CERN-LHCC-2017-009}, we derive
$P_{\text{conv}}(\eta)$ across the acceptance of the tracker from Eq.~\eqref{eq:conv_prob_t} for the Phase-1 tracker, which was the detector in operation during Run~3, and the Phase-2 tracker (Figs.~\ref{fig:phase1} and \ref{fig:phase2}).

\subsection{Displaced-vertex reconstruction}
\label{sec:vertex}
Since the ALP is long-lived (Sec.~\ref{sec:simulation}), its decay vertex is
displaced from the primary vertex. In Ref.~\cite{tracking-alps} this vertex was
reconstructed analytically from the conversion of two photons. The direction of each
converted photon is obtained from its conversion tracks, and the two directions are
extrapolated back to their common origin, which defines the ALP decay vertex. The
transverse distance to this vertex from the primary vertex gives the reconstructed
displacement $\tilde b$, and the resolution on $\tilde b$ is propagated from the
single-hit track resolution. Tagging the decay as displaced rejects the large prompt
backgrounds from hadronic states such as $\pi^0\to\gamma\gamma$. We follow this
procedure. Three experimental quantities control how well the reconstruction
performs. We take a conservative benchmark value for each and study how the projected
reach depends on these choices in Fig.~\ref{fig:parameter_space_variations}.

\paragraph{Track resolution.}
The resolution with which the direction of a track is measured sets the precision of
the displaced-vertex determination. For the Phase-2 IT, single-hit spatial resolutions
are $5$--$15\,\mu\mathrm{m}$ depending on pixel geometry and track incidence angle,
while beam-test measurements of the OT 2S strip modules yield
$\sigma \approx 27\,\mu\mathrm{m}$~\cite{CERN-LHCC-2017-009}, for an average single-hit
resolution of about $0.02\,\mathrm{mm}$. To be conservative we adopt twice this value,
$0.04\,\mathrm{mm}$, as the Phase-2 benchmark. For the Phase-1 tracker used in Run~3,
the pixel detector provides a hit resolution of about $10\,\mu\mathrm{m}$ in the
transverse and $20$--$40\,\mu\mathrm{m}$ in the longitudinal
coordinate~\cite{CMS:2014pgm,CMSTrackerGroup:2020edz}, and the strip sensors reach
$14$--$40\,\mu\mathrm{m}$ depending on strip pitch and track incidence
angle~\cite{CMS:2009xoa}. Taking an average single-hit resolution of about
$0.03\,\mathrm{mm}$ and applying the same factor of two, we adopt a conservative benchmark of
$0.06\,\mathrm{mm}$ for Run~3.

\paragraph{Vertex displacement requirement.}
This is the minimum $\tilde b$ required to tag a decay as displaced. In both Run~3 and Phase-2 strategies, we require
$\tilde b > 1\,\mathrm{cm}$ to remove the prompt hadronic backgrounds while
retaining the displaced signal.

\paragraph{Separation resolution.}
This is the minimum separation between the two photon tracks that can still be resolved
as distinct. We quantify it by a cut on the transverse separation of the two conversion
tracks at the outer wall of the tracker. To resolve two nearby tracks, their separation
must exceed the single-hit resolution by a comfortable margin. We take this margin to be
a factor of five, which gives $0.10\,\mathrm{mm}$ for Phase-2 and $0.15\,\mathrm{mm}$
for Run~3 from the average single-hit resolutions of $0.02\,\mathrm{mm}$ and
$0.03\,\mathrm{mm}$ quoted above.

\subsection{Data parking and data scouting in CMS}
\label{sec:datascouting}
Light ALPs are produced at the LHC in relatively low energy collisions, which result in soft
final state particles. Such particles present a challenge for the triggers of the LHC experiments. To keep the rates of SM backgrounds manageable, the LHC experiments select events online with thresholds on the kinematic features of the reconstructed particles. Events failing those selections are not recorded. Therefore, the data-acquisition strategies play a big role in achieving the sensitivity projected in this work. We focus on the CMS experiment whose trigger system offers dedicated ways to recover these otherwise discarded collisions.

The CMS experiment uses a two-tier trigger system. The L1T is built
from custom field-programmable gate arrays (FPGAs) and selects collisions within a
fixed latency of a few microseconds. In Run~3 it receives information from the
calorimeters and the muon system only. The Phase-2 upgrade adds tracks from the OT to the L1T inputs, making
track information available at the first trigger level for the first time
\cite{CERN-LHCC-2017-009, Bartz_2020, Tomalin:2287640}. With tracks, the
L1T can run particle-flow reconstruction, which combines the information from all
subdetectors to identify and reconstruct each particle in the event~\cite{pf}. The second tier is the high-level trigger (HLT), a processor farm that runs a version of the full reconstruction software optimized
to execute in a few milliseconds.

The CMS Collaboration developed two alternative trigger strategies, data parking and
data scouting~\cite{scouting-parking}, to recover collisions with soft particles that
standard triggers discard. The two address different limitations. Data parking tackles
the computing cost of the full, offline-quality reconstruction
of every recorded event that is normally run shortly after data-taking. It reduces this
cost by storing events in raw format and deferring their full reconstruction until
computing resources become available. This allows events to be recorded with relaxed
HLT selections. Data scouting instead addresses the limited trigger bandwidth, which is
set by both the data rate and the data size. It records only trigger-level objects, namely tracks, calorimeter
clusters, muons, and particle-flow candidates stored as four-vectors, without hit-level
or crystal-level detector information~\cite{L1DS-CHEP2025}. Since no raw data are stored, offline reconstruction of these
events is not possible. So far, data scouting has been operating only with the HLT event selection and reconstruction. The Phase-2 upgrade extends data scouting to the L1T~\cite{CERN-LHCC-2020-004}. The L1T will be able to perform particle-flow reconstruction and identify particles more
accurately, especially in the forward region, where it benefits from the finer space segmentation of the HGCAL. L1T data scouting will record collisions
with virtually no thresholds, increasing the acceptance for rare signals such as
VBF-produced ALPs with small couplings.

\subsection{Run~3 strategy with data parking}
\label{sec:run3}

During Run~3, CMS commissioned a set of VBF-specific parking triggers.
These were introduced part-way through the run, so the parked dataset corresponds to
$312\,\mathrm{fb}^{-1}$, out of the $355\,\mathrm{fb}^{-1}$ delivered throughout the
full Run~3~\cite{CMSLumiPublicResults}. For brevity we refer to this parked dataset as
Run~3 throughout the paper. The VBF parking triggers select collisions through the
kinematics of the two forward jets that characterize VBF production. Because the
trigger decision is based on the jets alone, no requirement is placed on the particles
produced in the fusion process. The soft photons from the ALP decay are therefore
recorded without any threshold. We use these
triggers to make the initial selection of signal events. Their combined thresholds are
given in Table~\ref{tab:vbf_trigger}.
\begin{table}[h]
\centering
\renewcommand{\arraystretch}{1.4}
\begin{tabular}{ll}
\hline\hline
\textbf{Variable} & \textbf{Threshold} \\
\hline
Leading jet $p_T$          & $> 105~\mathrm{GeV}$ \\
Sub-leading jet $p_T$      & $> 40~\mathrm{GeV}$  \\
Dijet invariant mass $m_{jj}$ & $> 720~\mathrm{GeV}$ \\
Dijet pseudorapidity gap $\Delta\eta_{jj}$ & $> 3.0$ \\
\hline\hline
\end{tabular}
\caption{Selection requirements of the CMS VBF data-parking trigger used in this analysis.}
\label{tab:vbf_trigger}
\end{table}

On this basis we select events that satisfy the following criteria. The
ALP must decay and its decay photons must convert within the tracker volume $|\eta| < 2.5$, so that the displaced
vertex can be measured. The diphoton separation $\Delta R_{\gamma\gamma}$ must be below the
size of a reconstructed photon at CMS, $\Delta R_{\gamma\gamma} < 0.3$, which
defines the merged-photon signal topology. The reconstructed transverse displacement of the decay vertex must satisfy
$\tilde b > 1\,\mathrm{cm}$, as in Sec.~\ref{sec:vertex}, which removes the prompt
hadronic backgrounds. The two conversion tracks must be separated by more than $0.15\,\mathrm{mm}$
at the outer wall of the tracker, so that they are resolved as distinct
(Sec.~\ref{sec:vertex}). Finally, the event must satisfy the VBF
parking trigger thresholds of Table~\ref{tab:vbf_trigger}. To tag photons as converted, we use the conversion probabilities derived for the Phase-1 tracker, the
detector in operation during Run~3 (Fig.~\ref{fig:phase1}), following the methodology
of Ref.~\cite{tracking-alps}. 

The displacement requirement removes the prompt backgrounds but not the long-lived
$K_L\to\gamma\gamma$.
Because the parked events retain the full detector readout, the merged diphoton
invariant mass can be reconstructed offline, and every $K_L$ decay reconstructs at
$m_{K_L}\simeq 0.5\GeV$. Away from this mass the $K_L$ is separated from the signal by
$m_{\gamma\gamma}$, and at $\ma\simeq m_{K_L}$ other kinematic variables provide the
remaining discrimination. We discuss the $K_L\to\gamma\gamma$ background and its suppression in detail in Sec.~\ref{sec:bkg}.

\subsection{Phase-2 strategy with data scouting}
\label{sec:phase2}
To reach ALPs with smaller couplings and thus smaller production cross sections, larger datasets are needed. The tenfold increase in integrated luminosity at the HL-LHC
(Sec.~\ref{sec:detector}) provides the rate. Recording the soft, rare signal produced
at small couplings additionally requires that it not be lost at the trigger.
As introduced in Sec.~\ref{sec:datascouting}, L1T data scouting
records collisions with virtually no thresholds, which increases acceptance for
this type of signal.

We propose a dedicated ``VBF Scouting Selection'' stream, defined from the hadronic
jets reconstructed in the L1T and consistent with VBF production. The jets in VBF originate from quarks, whereas the QCD background processes are dominated by gluon-initiated jets. The CMS Collaboration has developed AI techniques
\cite{particlenet}, including the Unified Particle Transformer (UParT) algorithm
\cite{upart} to select the
quark-initiated VBF jets. In a full CMS analysis, one can use a prototype L1T version of UParT
which provides the best quark-jet identification efficiency to date.

\begin{figure}
    \centering
    \begin{subfigure}{0.49\linewidth}
        \includegraphics[width=\linewidth]{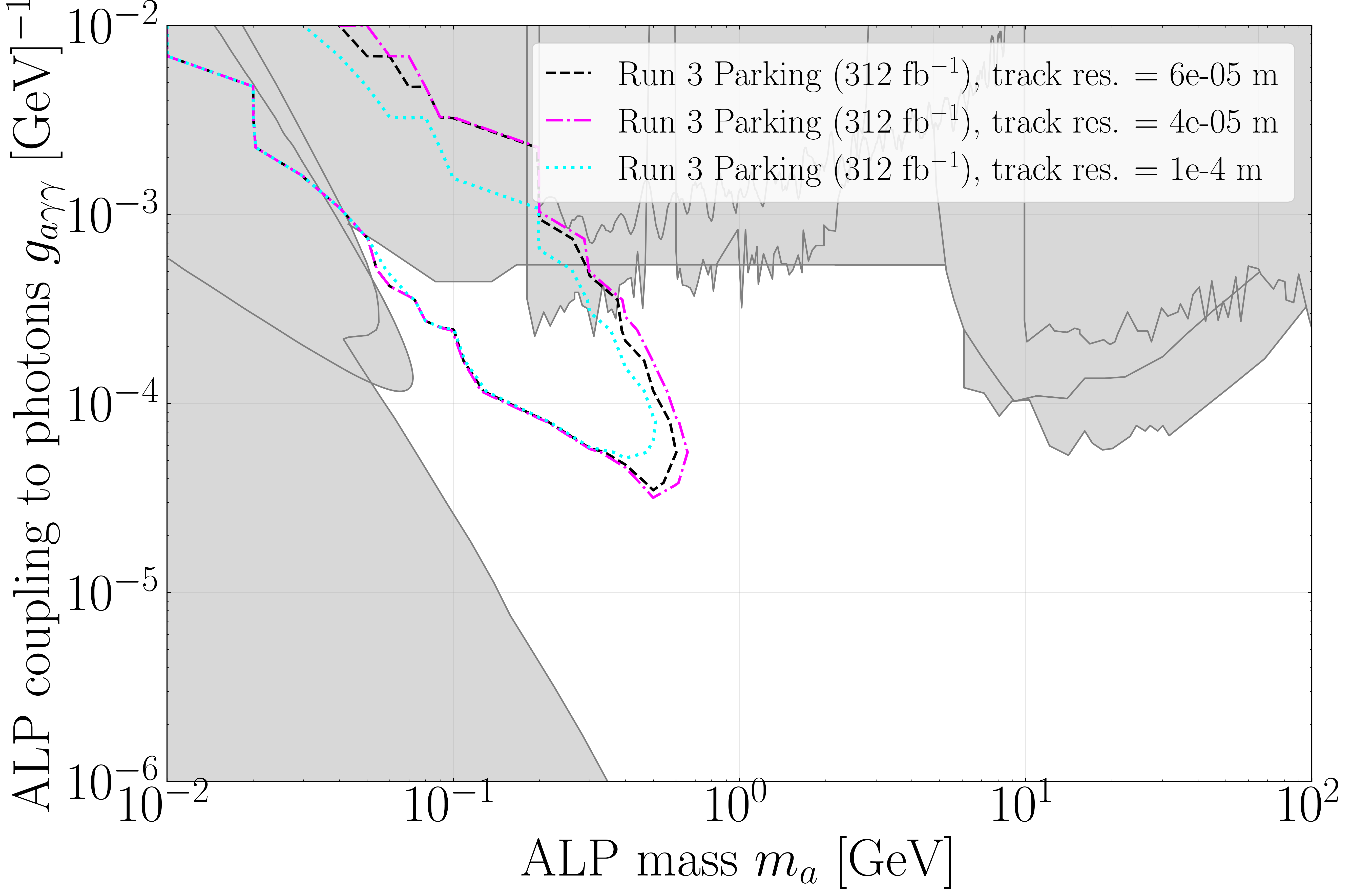}
        \caption{}
        \label{fig:exp-track-res}
    \end{subfigure}%
    \hfill
    \begin{subfigure}{0.49\linewidth}
        \includegraphics[width=\linewidth]{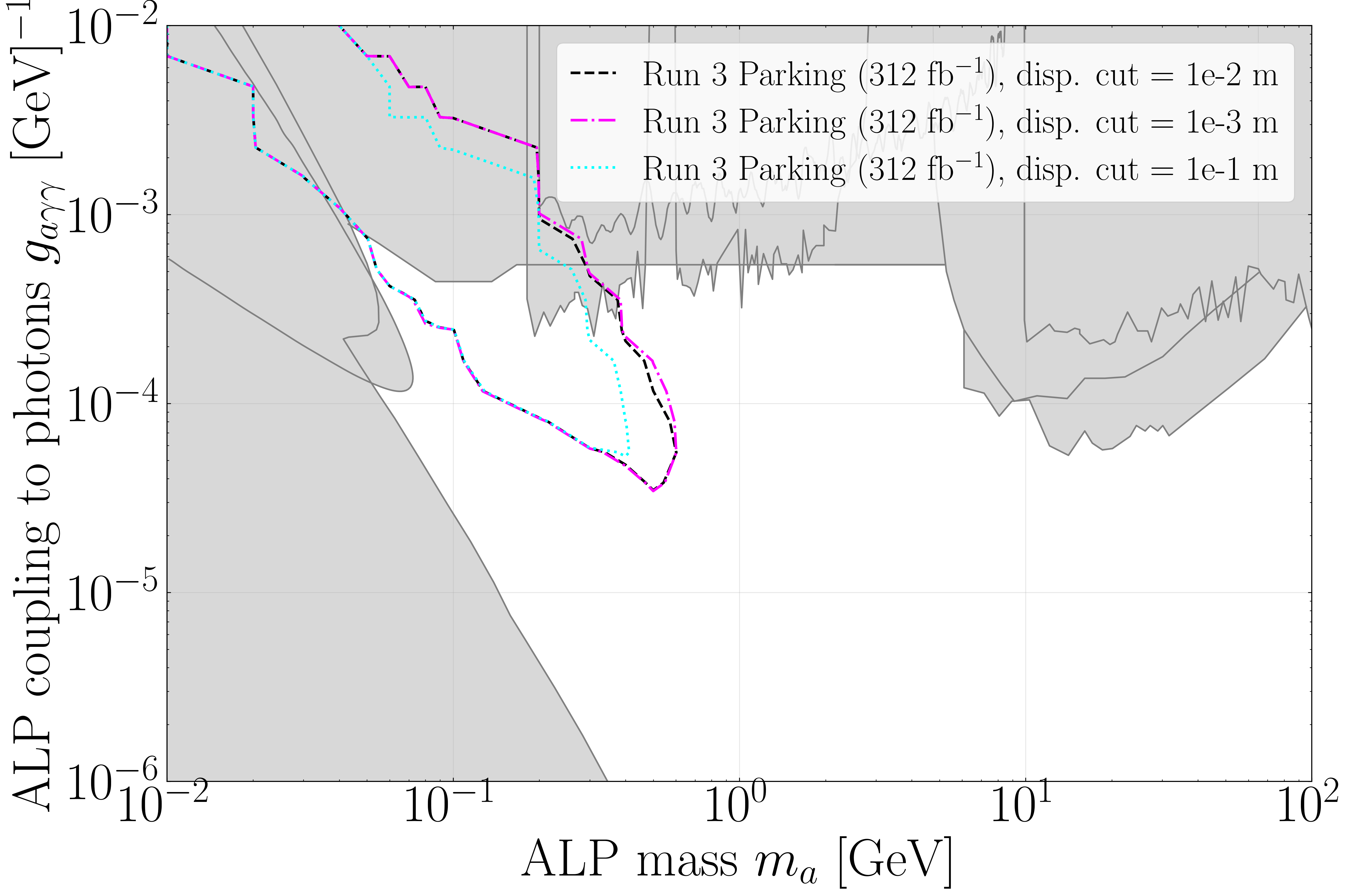}
        \caption{}
        \label{fig:exp-disp-cut}
    \end{subfigure}
    \hfill
    \begin{subfigure}{0.49\linewidth}
        \includegraphics[width=\linewidth]{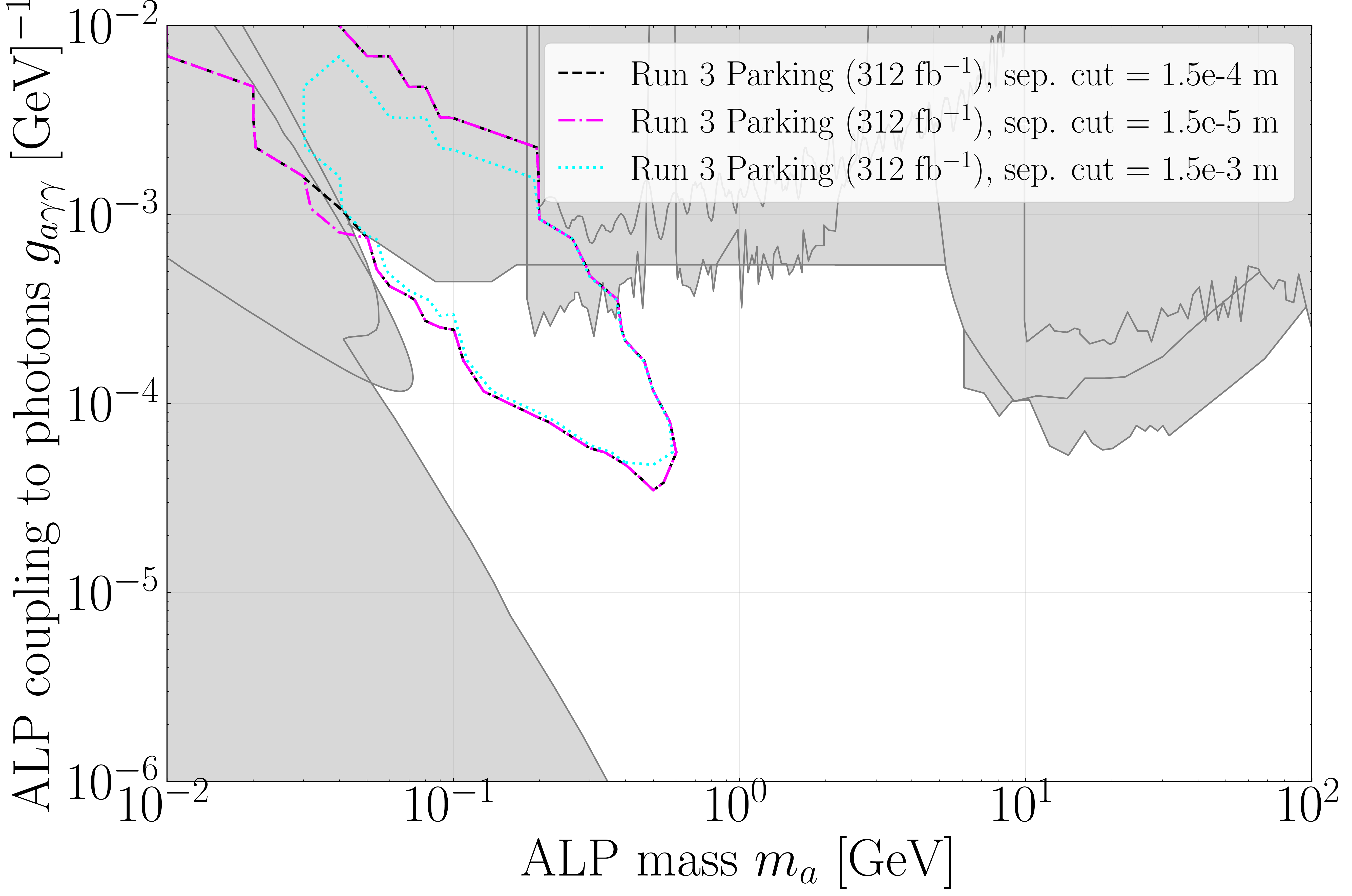}
        \caption{}
        \label{fig:exp-sep-cut}
    \end{subfigure}    
    \begin{subfigure}{0.49\linewidth}
        \includegraphics[width=\linewidth]{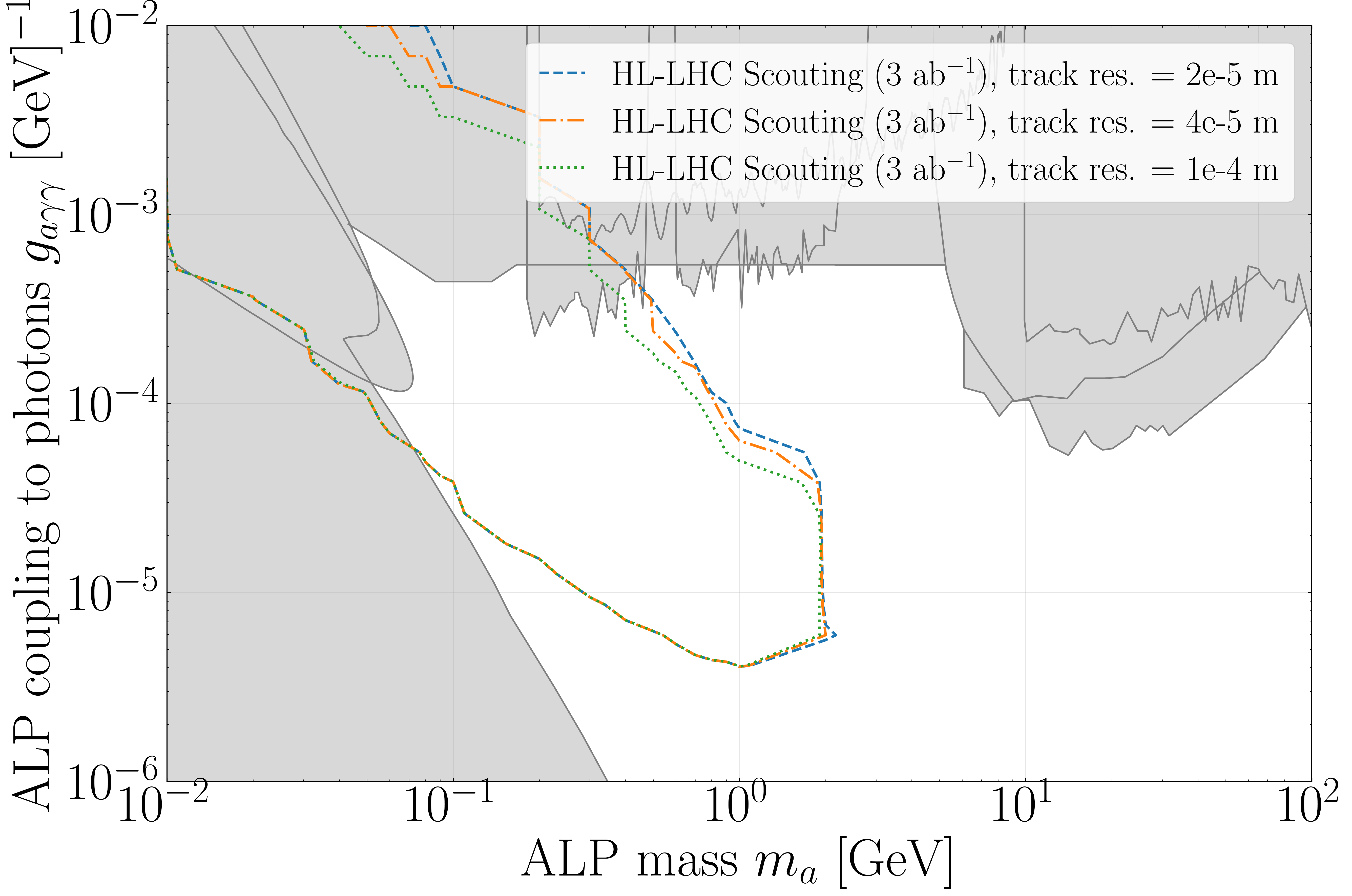}
        \caption{}
        \label{fig:run3-disp}
    \end{subfigure}
    \begin{subfigure}{0.49\linewidth}
        \includegraphics[width=\linewidth]{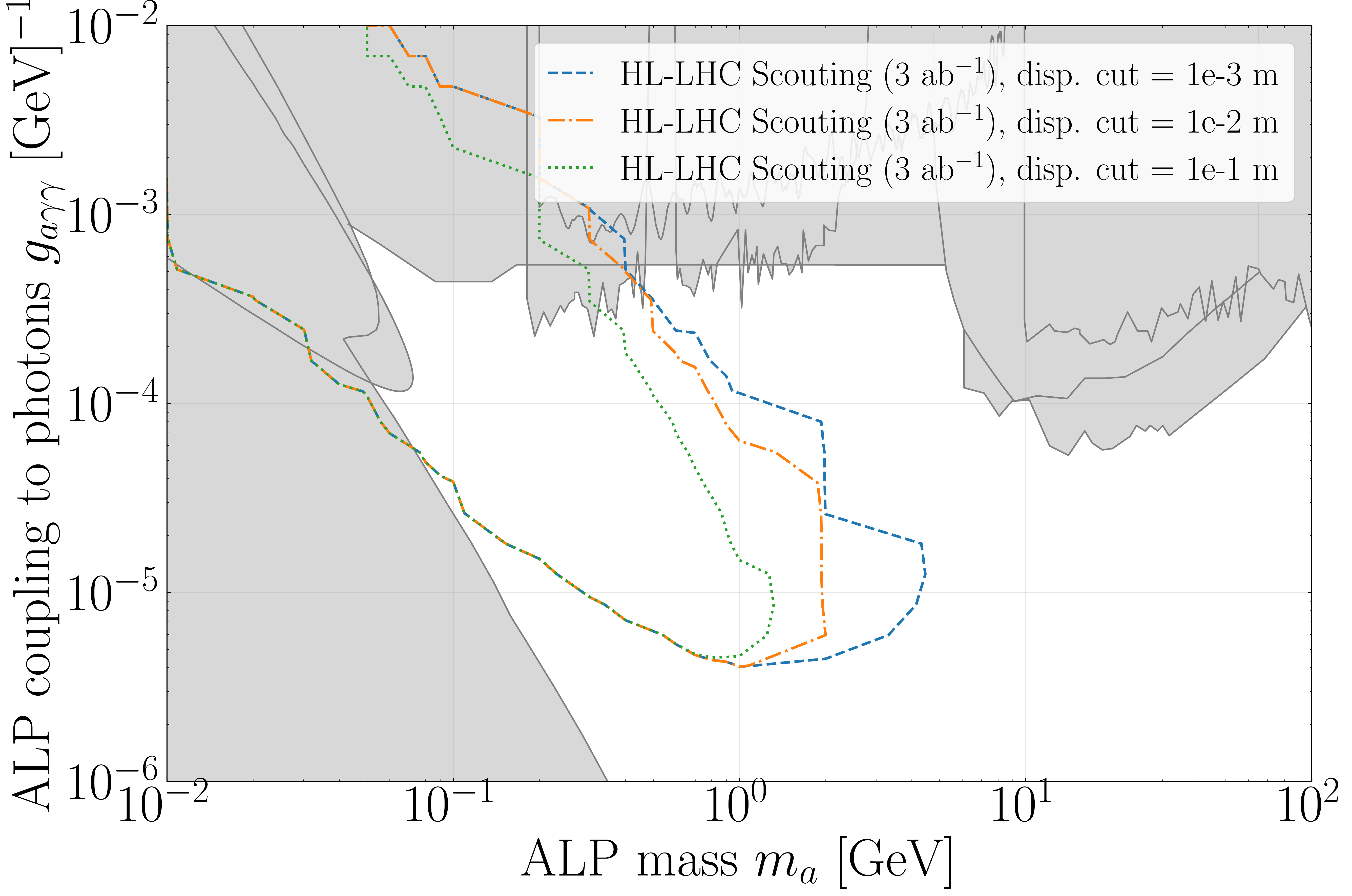}
        \caption{}
        \label{fig:run3-trackres}
    \end{subfigure}
    \begin{subfigure}{0.49\linewidth}
        \includegraphics[width=\linewidth]{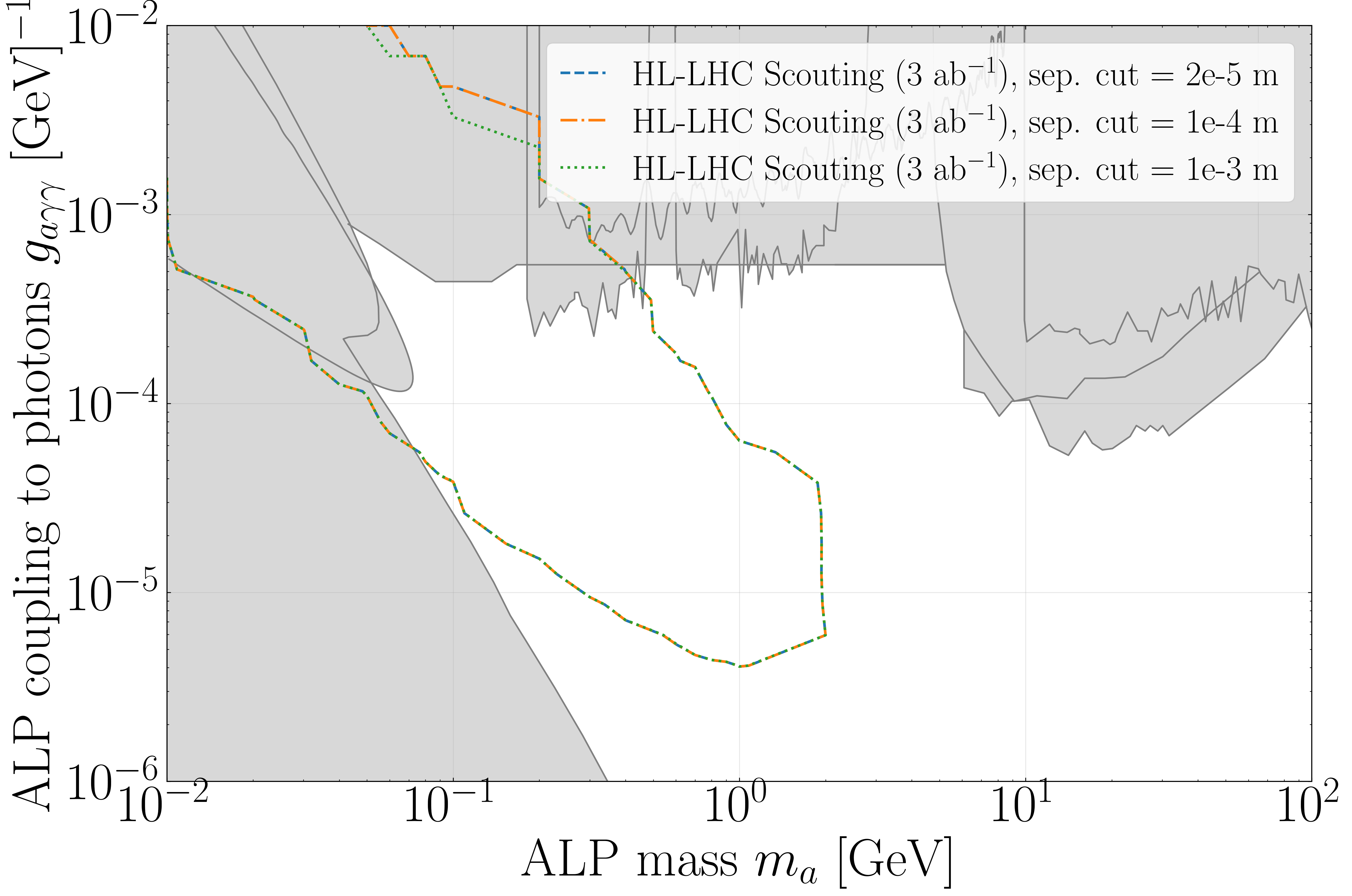}
        \caption{}
        \label{fig:run3-trackres}
    \end{subfigure}    

    \caption{Effect of varying the track resolution, displacement requirement, and
    separation resolution (Sec.~\ref{sec:vertex}) on the projected reach in the ALP
    mass versus photon coupling plane. The gray area is excluded by other experiments.}
    \label{fig:parameter_space_variations}
\end{figure}

For the data-scouting projection, we select events in which the ALP decays within the
tracker volume, both photons fall within the HGCAL acceptance of $1.4<|\eta|<3.0$,
where the conversion probability plateaus at $\approx40\%$ (Fig.~\ref{fig:phase2}),
both photons convert within the tracker, and the photon separation satisfies
$\Delta R_{\gamma\gamma}<0.3$. The decay is tagged as displaced through the
$\tilde b > 1\,\mathrm{cm}$ requirement of Sec.~\ref{sec:vertex}. Finally, the transverse separation of the two conversion tracks at the outer tracker wall must exceed 0.1 mm, so that they are resolved as distinct.

As in Run~3, the prompt backgrounds are rejected through the displaced-vertex
reconstruction of Sec.~\ref{sec:vertex}, and the long-lived $K_L\to\gamma\gamma$
survives this requirement. The difference lies in what remains to separate the $K_L$
from the signal. Since the scouting stream stores no raw data, the offline
merged-diphoton mass reconstruction available in Run~3 cannot be performed. The
discrimination must instead come from the diphoton $p_T$, which is much harder for the
signal than for the $K_L$, from the quantities accessible through the conversion
tracks, such as the individual photon momenta and their opening angle, and from the
hadronic activity in the event. The photon
momenta and opening angle also carry information on the diphoton mass. Building a mass
estimator from trigger-level track parameters and calorimeter energies, without
crystal-level shower information, is a hard problem, but it would recover much of the
discriminating power lost with the raw data. The time remaining before HL-LHC data
taking leaves room for such a tool to be developed. In the following section, we discuss the $K_L$ simulation and the kinematic separation in greater detail.

\section{Backgrounds}
\label{sec:bkg}
We treat only the backgrounds that produce a genuine collimated photon pair at Monte Carlo truth level. Backgrounds that
enter the signal region solely through a jet$\to\gamma$ fake, an $e\to\gamma$
fake, or a misreconstruction require a full detector simulation and data-driven
control samples. They lie outside the scope of this phenomenology-level study and are
deferred to a future experimental analysis. 

The backgrounds can be grouped according to whether they are prompt. Prompt backgrounds are neutral mesons $\pi^0,\eta,\eta'\to\gamma\gamma$ and the SM prompt diphoton continuum. Ref.~\cite{tracking-alps} showed that imposing a requirement on the reconstructed displaced vertex removes essentially the entire $\pi^0$ contribution, and we adopt that
conclusion here. The same argument applies to the remaining prompt sources. The
light mesons $\eta$ and $\eta'$ have laboratory decay lengths that are microscopic
on collider scales, and, together with the prompt diphoton continuum, are  
removed by the $\tilde b > 1\,\mathrm{cm}$ requirement, where
$\tilde b$ is the reconstructed transverse displacement of the diphoton vertex from
the primary vertex. We apply this requirement in both the Run~3 and Phase-2 strategies described in this study.

The second group of backgrounds is long-lived. This contribution, dominated by the long-lived neutral kaon $K_L$, survives the displacement cut and therefore requires different treatment.
The relaxed trigger thresholds considered here compared to those in \cite{tracking-alps} introduce a significant portion of this long-lived background into the analysis\footnote{We furthermore find that~\cite{tracking-alps} likely underestimated the $K_L$ background in the low-$p_T$ region due to the extrapolation procedure employed.}.
\subsection{$K_L\to\gamma\gamma$}
\label{sec:kl}
The $K_L$ has a lifetime of $\tau_{K_L}\simeq5\times10^{-8}$~s
($c\tau\simeq15.3$~m) and a branching ratio to two photons $\mathrm{Br}(K_L\to\gamma\gamma)\simeq5.47\times10^{-4}$ \cite{pdg}. A $K_L$ decaying inside the tracker volume produces a collimated and
displaced diphoton vertex. The discriminating power against the signal therefore comes not primarily from the displacement but from other kinematic variables. In what follows we discuss the simulation of the $K_L$ samples and then these variables.

\subsubsection{Simulation}
We use the \textsc{Foresee} package~\cite{foresee} and the \textsc{Pythia}
generator~\cite{pythia}. \textsc{Foresee} provides the inclusive momentum spectra
of forward hadrons including $K_L$. This spectrum is used for initial rate
estimates. For the suggested Run~3 data parking strategy, the full event, in particular the
 jet kinematics, is needed to impose the VBF trigger selection. \textsc{Pythia}
provides this information. The \textsc{Foresee} and \textsc{Pythia} predictions agree
on the shape of the inclusive $K_L$ momentum spectrum and on the integrated forward yield
at the few-percent level.

We produce two complementary samples with \textsc{Pythia}. We generate a soft-QCD
sample of $2\times10^{6}$ minimum-bias events to simulate the full inelastic $pp$
collision ($\sigma_{\mathrm{inel}}=78.6$~mb, consistent with the measured
values~\cite{ATLAS:2016ygv, TOTEM:2017asr}), which gives us the unbiased inclusive
$K_L$ yield and kinematics. Collisions capable of satisfying the VBF selections given in Table~\ref{tab:vbf_trigger} constitute only $\mathcal{O}(10^{-6})$ of the
inelastic rate, far too rare to sample efficiently from minimum bias. We therefore
generate a dedicated hard-QCD sample of $2\times10^{7}$ dijet events
($\hat{p}_{T}>80$~GeV, $\sigma = 3.6\times10^{-3}$~mb) to populate this tail
directly. We use the hard-QCD sample to fix the VBF-passing cross section and the
kaon kinematics in VBF-triggered events, and the soft-QCD sample to set the inclusive rate. We use the hard-QCD sample for the background plots corresponding to Run~3, and the soft-QCD sample for those corresponding to Phase-2. 

The soft-QCD sample yields an average multiplicity of
$\langle n_{K_L}\rangle = 4.1$ per inelastic collision. Combined with
$\sigma_{\mathrm{inel}}$ and the integrated luminosity, this corresponds to about
$1.0\times10^{17}$ $K_L$ produced in 312~\fbinv\ ($9.6\times10^{17}$ in
3000~\fbinv), reduced by the diphoton branching ratio to $5.4\times10^{13}$
($5.2\times10^{14}$).

\subsubsection{Kinematic discrimination}
The diphoton invariant mass is the primary discriminant against
$K_L\to\gamma\gamma$. Every $K_L$ decay reconstructs at
$m_{\gamma\gamma}\simeq m_{K_L}\simeq 0.498\GeV$, smeared only by the diphoton mass
resolution $\sigma_{m_{\gamma\gamma}}$. For any ALP hypothesis with $\ma$ outside a
window of width $\sim\sigma_{m_{\gamma\gamma}}$ around $m_{K_L}$, the $K_L$
background does not populate the signal region and the search is approximately
background-free under the assumptions stated below.
The one exception is $\ma\approx m_{K_L}$ where the two resonances overlap. In
that bin the sensitivity is set by the $K_L$ yield within the mass resolution
rather than by a background-free assumption. Reconstructing the diphoton invariant mass for merged, collimated photon pairs is non-trivial and requires dedicated
techniques. A recent study explored Transformer architectures for this purpose
using a simplified simulation of the CMS ECAL. The ClusTEX algorithm~\cite{clustex}
performs reconstruction in a single stage by dynamically forming a graph around a
seed candidate, and reconstructs a sharp peak around the neutral pion mass with a
resolution of $\sim 14\%$. We therefore operate under the assumption that a preliminary version of this
reconstruction method is available for Run~3 within CMS, which allows us to use the
invariant mass as a discriminant. We emphasize that adapting ClusTEX, or developing a
dedicated algorithm for our topology, is essential to achieve the Run~3 sensitivity
projected in this study. Incorporating track information into such an algorithm would
further improve performance by identifying converted photons and differentiating them
from single photons. This would be particularly beneficial for low-mass ($<100$~MeV)
ALPs, where the ECAL granularity alone is insufficient due to the high collimation of
the photon pair. In the eventual CMS analysis of the Run~3 data, the search would take
the form of a hunt for a narrow peak in the diphoton invariant mass over a smooth
continuum modeled from data, with the tracks and the kinematic variables discussed
below providing additional discrimination. For Phase-2, the trigger-level information
in the L1T data scouting stream is too limited for this type of mass reconstruction. Our
Phase-2 projections therefore rely on the tracks and the kinematic variables alone. 
Developing an invariant-mass reconstruction from the information available in the
scouting stream would add the most powerful discriminant to the Phase-2 search, and
we consider it a highly valuable avenue for future work.

\begin{figure}
    \centering
    \begin{subfigure}[t]{0.48\textwidth}
        \centering
        \includegraphics[width=\textwidth]{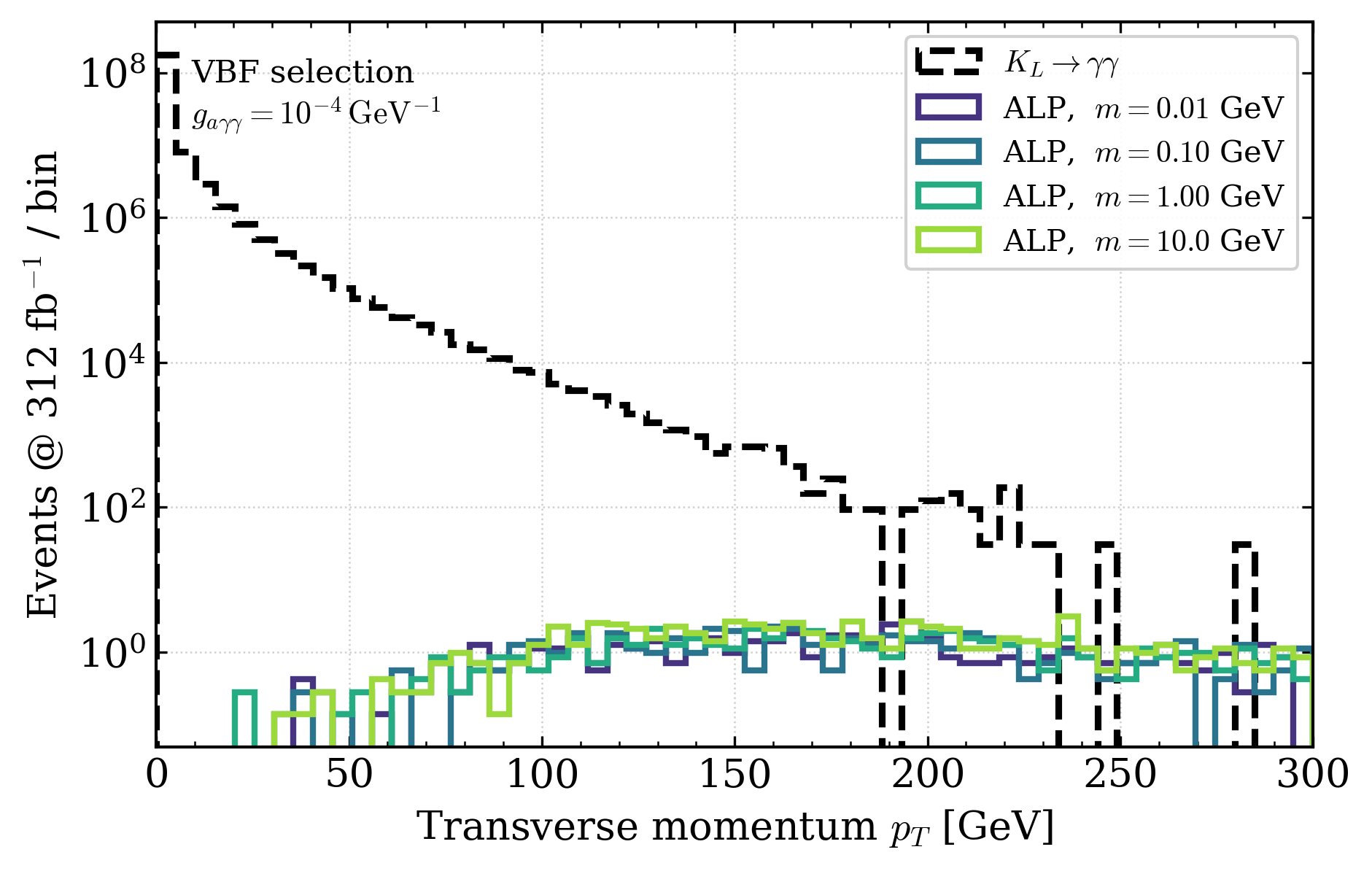}
        \caption{VBF-triggered (Run~3 parking), $312$~\fbinv.}
        \label{fig:pt_vbf}
    \end{subfigure}
    \hfill
    \begin{subfigure}[t]{0.48\textwidth}
        \centering
        \includegraphics[width=\textwidth]{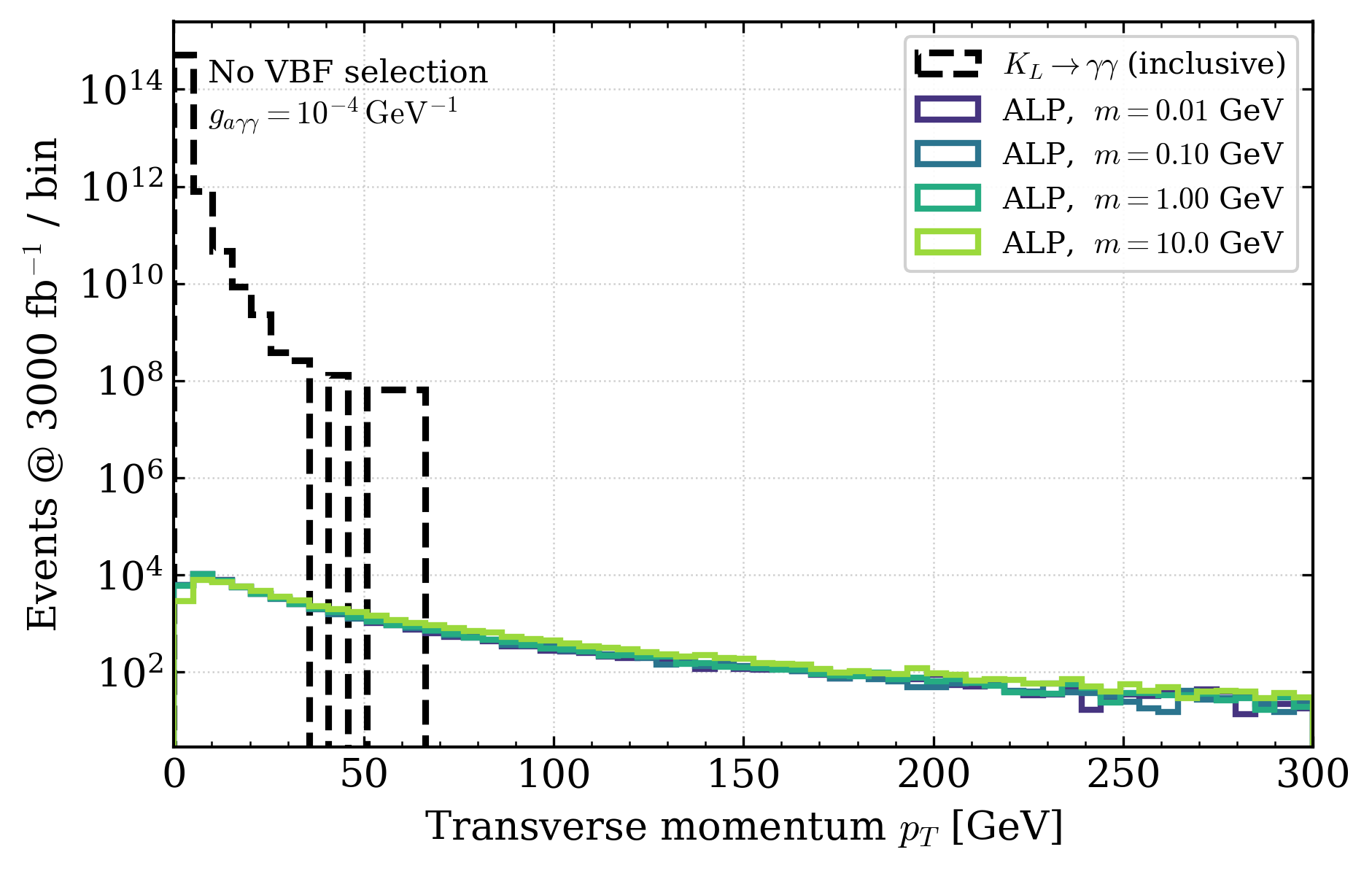}
        \caption{No trigger requirement (HL-LHC scouting), $3000$~\fbinv.}
        \label{fig:pt_incl}
    \end{subfigure}

    \vspace{1em}

    \begin{subfigure}[t]{0.48\textwidth}
        \centering
        \includegraphics[width=\textwidth]{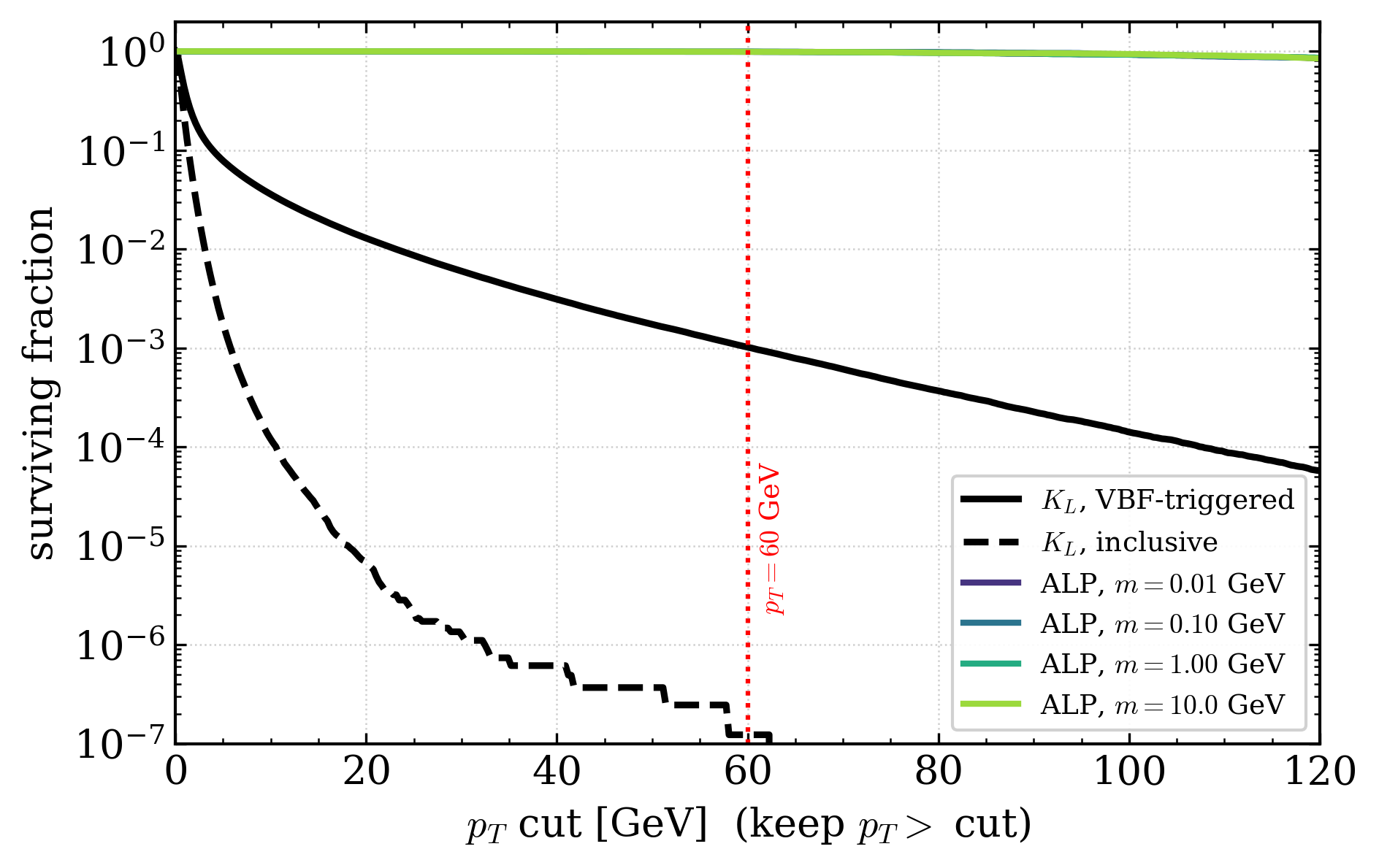}
        \caption{Survival fraction versus $p_T$ threshold.}
        \label{fig:pt_eff}
    \end{subfigure}
    \caption{Diphoton system $p_T$ for the ALP signal (colored, four masses at
$\gagg=10^{-4}\,\mathrm{GeV}^{-1}$) and the $K_L\to\gamma\gamma$ background (black),
normalized to physical yields. The $K_L$ population falls off more steeply than the ALP population in both cases. As shown in (c), a $60$~GeV requirement on the
parent ALP or $K_L$ keeps $\gtrsim99\%$ of the signal but only
$<\mathcal{O}(10^{-4})$ of the VBF-triggered and $<\mathcal{O}(10^{-7})$ of the
inclusive kaons.}
    \label{fig:pt_all}
\end{figure}

We also compare the predicted diphoton $p_T$ distributions of the signal and the $K_L$ background. The \kl mesons are often the product of soft hadronization and are overwhelmingly soft. The soft-QCD sample has a median $p_T\simeq0.5$~GeV, and even in the hard-QCD events the kaon spectrum falls quite steeply. The ALP spectrum extends to much higher $p_T$
in both populations (Figs.~\ref{fig:pt_vbf} and \ref{fig:pt_incl}). A requirement of $p_T>60$~GeV in the Run~3
population would retain essentially all of the VBF-selected signal while rejecting
all but $<\mathcal{O}(10^{-4})$ of the triggered kaons. In the scouting population
the same requirement rejects all but $<\mathcal{O}(10^{-7})$ of the inclusive
kaons (Fig.~\ref{fig:pt_eff}). These rejection factors are set by the finite size of the simulated samples. No
simulated $K_L$ populates the region above the threshold, so the quoted values reflect
the resolution of our samples and are not measurements of the surviving fraction. The
$K_L$ spectrum is steeply decreasing but does not vanish at high $p_T$, and given the very
large number of kaons produced, a small surviving fraction still corresponds to a
sizable absolute yield. Quantifying it requires either far larger samples or an
estimate of the tail directly from data, which is beyond the scope of this study. For the sensitivity projection in Fig.~\ref{fig:sensitivity} we do not impose a threshold on the $p_T$.  A fixed cut discards the shape information contained in the
full $p_T$ spectrum. In a complete CMS analysis the diphoton $p_T$ would instead
enter as a continuous discriminant, either as an input to a multivariate classifier
or as a fit variable in a likelihood built jointly with $m_{\gamma\gamma}$ and the
decay displacement. The residual $K_L$ contribution would then be constrained
directly from data using standard data-driven background estimation. We therefore
treat the $p_T$ separation shown here as a conservative lower bound on the achievable
rejection, rather than the selection actually used.

\begin{figure}[htbp]
    \centering
    \begin{subfigure}[t]{0.48\textwidth}
        \centering
        \includegraphics[width=\textwidth]{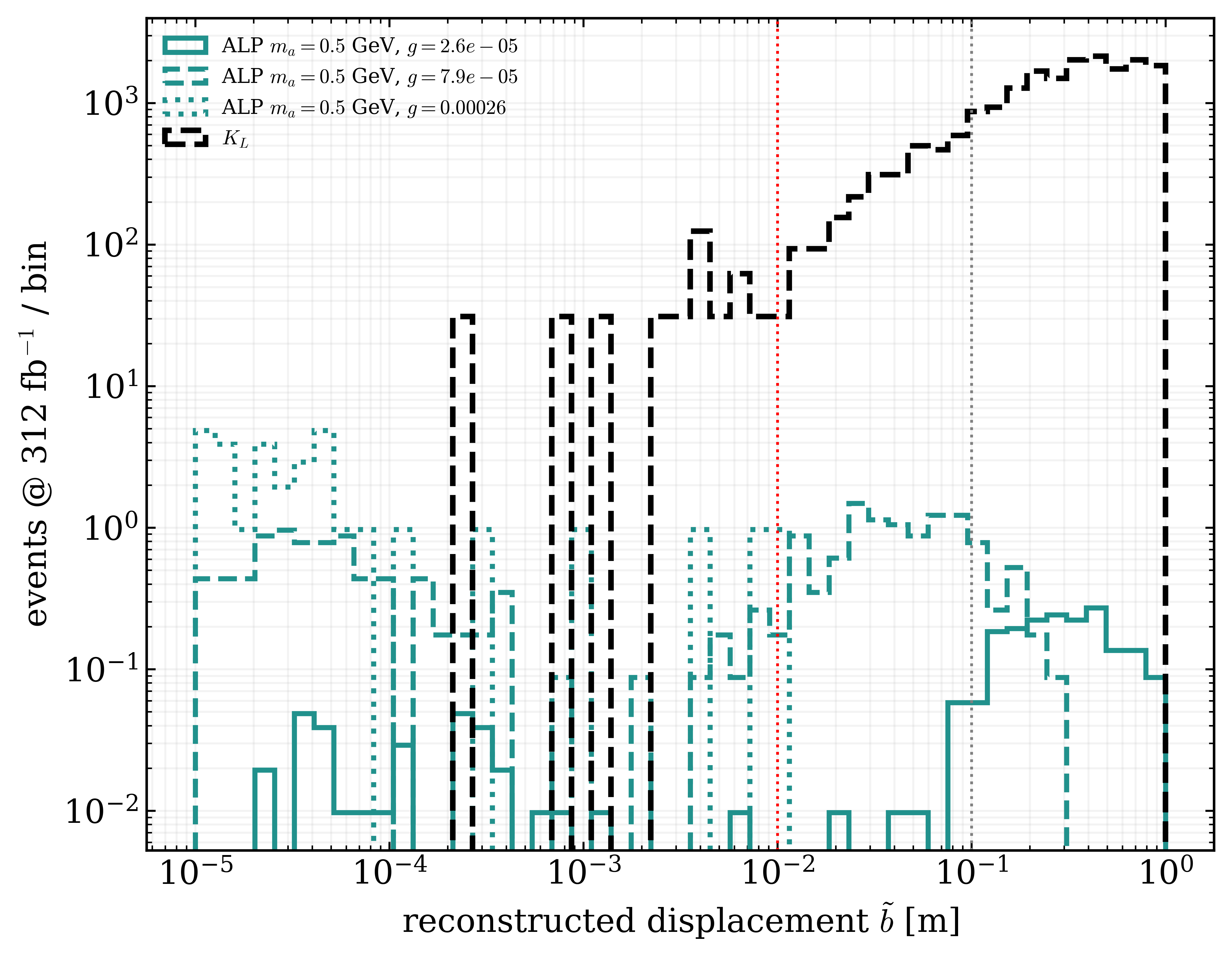}
        \caption{Run~3 (VBF-triggered parking), physical yields at $312$~\fbinv.}
        \label{fig:disp_run3_yield}
    \end{subfigure}
    \hfill
    \begin{subfigure}[t]{0.48\textwidth}
        \centering
        \includegraphics[width=\textwidth]{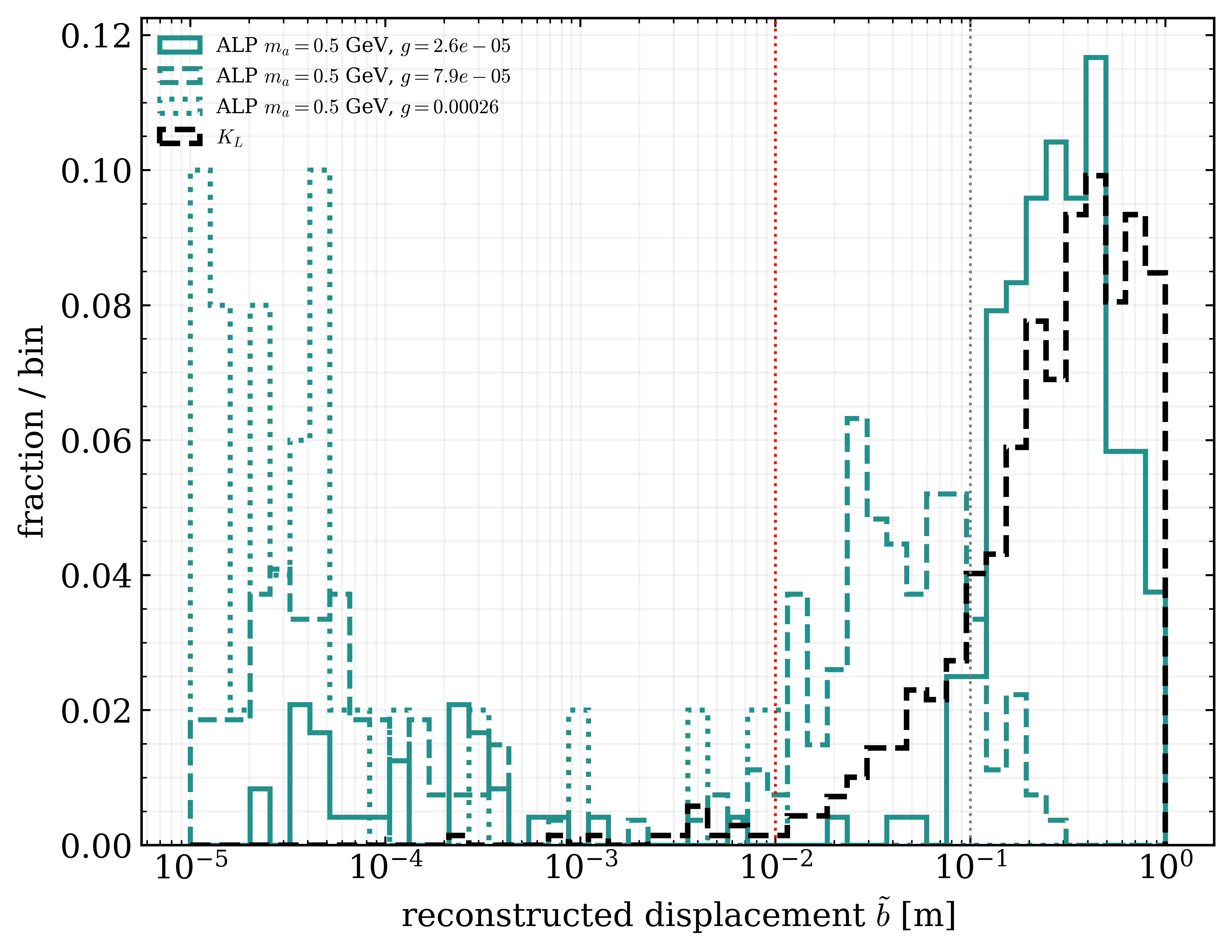}
        \caption{Run~3, normalized shape.}
        \label{fig:disp_run3_shape}
    \end{subfigure}

    \vspace{1em}

    \begin{subfigure}[t]{0.48\textwidth}
        \centering
        \includegraphics[width=\textwidth]{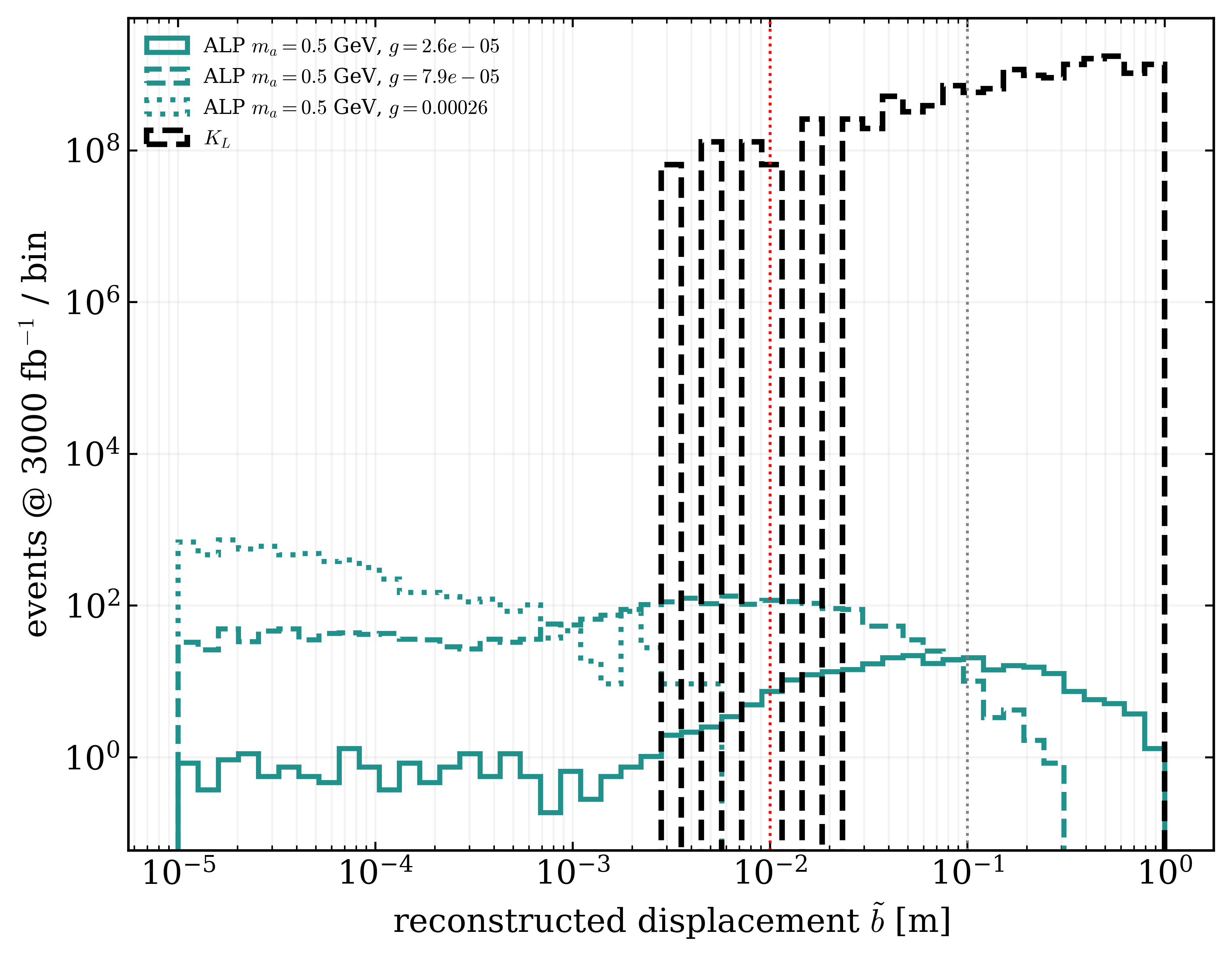}
        \caption{HL-LHC scouting (Phase-2), physical yields at $3000$~\fbinv.}
        \label{fig:disp_phase2_yield}
    \end{subfigure}
    \hfill
    \begin{subfigure}[t]{0.48\textwidth}
        \centering
        \includegraphics[width=\textwidth]{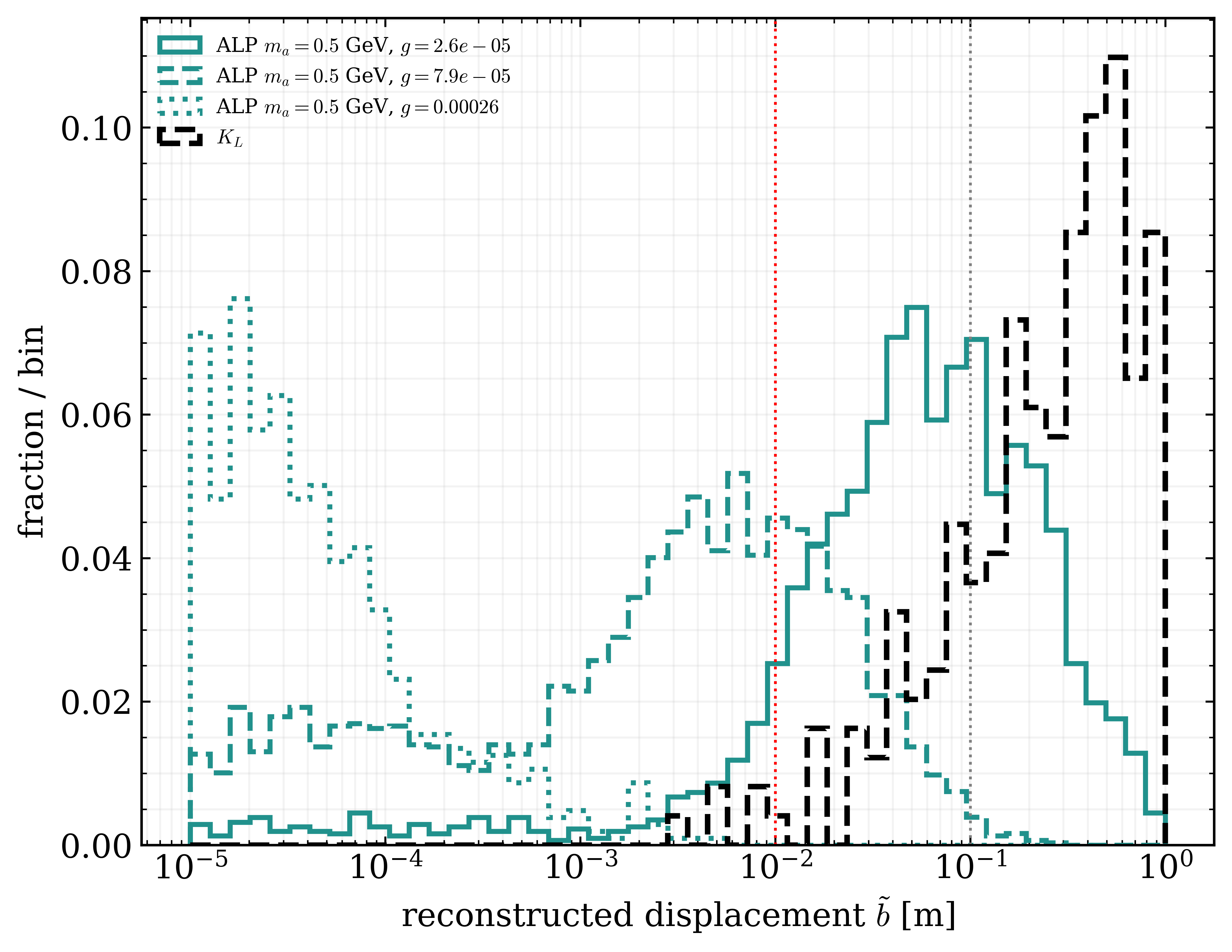}
        \caption{HL-LHC scouting (Phase-2), normalized shape.}
        \label{fig:disp_phase2_shape}
    \end{subfigure}
    \caption{Reconstructed displacement $\tilde b$ for the ALP signal (colored,
$\ma = 0.5\GeV \approx m_{K_L}$, three couplings) and the $K_L\to\gamma\gamma$
background (black), with no kinematic selections applied. Top row: the Run~3 population
(VBF-triggered, hard-QCD $K_L$) at $312$~\fbinv. Bottom row: the HL-LHC scouting
population (inclusive, soft-QCD $K_L$) at $3000$~\fbinv. Left column: physical yields.
Right column: normalized shape. The red dotted line marks the $\tilde b > 1\,$cm
displacement requirement. The $K_L$ dominates the signal at all displacements. The
depletion of the $K_L$ at small $\tilde b$ reflects the finite simulated statistics
rather than a genuine absence of background.}
    \label{fig:disp_all}
\end{figure}

Figure ~\ref{fig:disp_all} shows the reconstructed displacement for the $K_L$ and for
the ALP at $\ma = 0.5\GeV \approx m_{K_L}$, the mass at which the invariant mass gives
no separation. Three couplings are shown, spanning short to long ALP lifetimes, with
no $p_T$ selection applied. The top row is the Run~3 population (VBF-triggered,
hard-QCD $K_L$) and the bottom row is the Phase-2 population (inclusive, soft-QCD
$K_L$). The normalized distributions (right column) show that the displacement
separates signal from background in shape only at short lifetime. At large coupling
the ALP decays early and peaks at small $\tilde b$, while the $K_L$ rises toward the
tracker edge near $1\,\mathrm{m}$. At small coupling the ALP is long-lived and its
shape coincides with the $K_L$.

At physical yields (left column) the $K_L$ exceeds the signal across the full
displacement range. In the Run~3 population the VBF trigger suppresses the $K_L$ to
$\mathcal{O}(10^{3})$ events per bin, still far above the signal. In the Phase-2
population no such suppression is applied, and the inclusive $K_L$ reaches
$\mathcal{O}(10^{8})$ events per bin. Displacement alone therefore does not isolate
the signal at this mass. The apparent absence of $K_L$ at small $\tilde b$ in the
yield panels reflects the finite simulated statistics rather than a genuine absence
of background. The physical $K_L$ yield is very large, a small fraction decays early,
and our samples do not populate that regime. We therefore do not treat the
small-$\tilde b$ region as background-free.

These distributions do not include the $p_T$ selection. Applying it suppresses the
$K_L$ by $\mathcal{O}(10^{-4})$ in the Run~3 population and $\mathcal{O}(10^{-7})$ in
the Phase-2 population, which is what brings the residual $K_L$ near
$\ma\simeq m_{K_L}$ down toward the signal level. Away from the kaon mass the invariant mass removes the $K_L$ entirely, and the
displacement instead serves to reject the prompt backgrounds. 

The signal selections listed in Sec.~\ref{sec:strategy} suppress part of the
background as well. Two further properties of the events
provide additional suppression. First, the jets in VBF originate from quarks, whereas
the QCD processes that produce the $K_L$ are dominated by gluon-initiated jets.
Discriminating quark jets from gluon jets can therefore remove a large fraction of
this background. As mentioned in Sec.~\ref{sec:phase2}, this can be achieved with the
UParT algorithm. Second, a $K_L$ is produced inside a jet, so the diphoton candidate
is accompanied by nearby charged and neutral particles, whereas the ALP recoils
against the two forward VBF jets and is otherwise isolated. Requiring the diphoton
candidate to be isolated therefore rejects the $K_L$ while retaining the signal.
Neither of these is included in the yields quoted here, so the $K_L$ yields in
Table~\ref{tab:yields} are conservative. The table gives the expected cross section,
selection efficiency, and yields for the signal and the $K_L$ in both Run~3 and
Phase-2.

How these features combine depends on the acquisition stream. In the Run~3 data
parking stream, the full event is recorded, so the invariant mass is the leading
discriminant. Combined with the displacement in a two-dimensional
$(m_{\gamma\gamma},\tilde b)$ fit, the mass separates everywhere except the narrow
window $|\ma - m_{K_L}| \lesssim \sigma_{m_{\gamma\gamma}}$, where the diphoton $p_T$
and the displacement provide the remaining suppression. In Phase-2 no invariant-mass
reconstruction is available, so the suppression relies instead on the diphoton $p_T$,
the displacement, and the hadronic environment. The $p_T$ separation is especially
effective here, since the inclusive $K_L$ spectrum is very soft and a requirement of
$p_T>60$~GeV already removes all but $\mathcal{O}(10^{-7})$ of the inclusive kaons.
In a full analysis the per-photon observables accessible from the converted-photon
tracks, such as the individual photon momenta and the diphoton opening angle, could
be added to further suppress the residual $K_L$, and a mass estimator built from them
at the trigger level would be the most powerful addition. We treat the search as
background-free in both strategies, and the sensitivity projections in
Sec.~\ref{sec:results} are computed under this assumption. In the scouting stream
this assumption requires further work along the lines described above.

\begin{table}
\caption{\label{tab:yields}Sample cross section $\sigma$, selection efficiency
$\varepsilon$, and expected yield $N$ for the $K_L\to\gamma\gamma$ background and
signal benchmarks, for Run~3 parking ($312~\mathrm{fb}^{-1}$) and Phase-2
scouting ($3~\mathrm{ab}^{-1}$). ALP masses $m_a$ are in GeV, couplings $g$ are in $\mathrm{TeV}^{-1}$. The
Run~3 $K_L$ sample is hard-QCD ($\hat{p}_T^{\mathrm{min}}=80$~GeV), the Phase-2
sample is inclusive.}
\footnotesize
\setlength{\tabcolsep}{3pt}
\centering
\begin{tabular}{lccc}
\hline\hline
Process & $\sigma$ [pb] & $\varepsilon$ & $N$ \\
\hline
\multicolumn{4}{c}{Run~3 parking, $312~\mathrm{fb}^{-1}$}\\
\hline
$K_L\to\gamma\gamma$              & $3.64\times10^{6}$  & $3.62\times10^{-8}$ & $4.11\times10^{4}$ \\
$m_a=0.02$, $g=10$                & $1.80\times10^{2}$  & $3.50\times10^{-4}$ & $1.97\times10^{4}$ \\
$m_a=0.1$, $g=1.08$               & $2.09$              & $1.53\times10^{-3}$ & $9.94\times10^{2}$ \\
$m_a=1$, $g=2.63\times10^{-2}$    & $1.24\times10^{-3}$ & $1.70\times10^{-3}$ & $<1$ \\
$m_a=10$, $g=3.05\times10^{-4}$   & $1.67\times10^{-7}$ & $3.15\times10^{-3}$ & $<10^{-3}$ \\
\hline
\multicolumn{4}{c}{Phase-2 scouting, $3~\mathrm{ab}^{-1}$}\\
\hline
$K_L\to\gamma\gamma$              & $7.85\times10^{10}$ & $9.93\times10^{-8}$ & $2.34\times10^{10}$ \\
$m_a=0.02$, $g=10$                & $1.80\times10^{2}$  & $3.77\times10^{-2}$ & $2.04\times10^{7}$ \\
$m_a=0.1$, $g=1.08$               & $2.09$              & $3.35\times10^{-2}$ & $2.10\times10^{5}$ \\
$m_a=1$, $g=1.81\times10^{-2}$    & $5.91\times10^{-4}$ & $2.55\times10^{-2}$ & $45$ \\
$m_a=10$, $g=3.05\times10^{-4}$   & $1.67\times10^{-7}$ & $6.93\times10^{-3}$ & $<10^{-2}$ \\
\hline\hline
\end{tabular}
\end{table}

\section{Results}
\label{sec:results}
\begin{figure}

    \centering
    \includegraphics[width=0.85\linewidth]{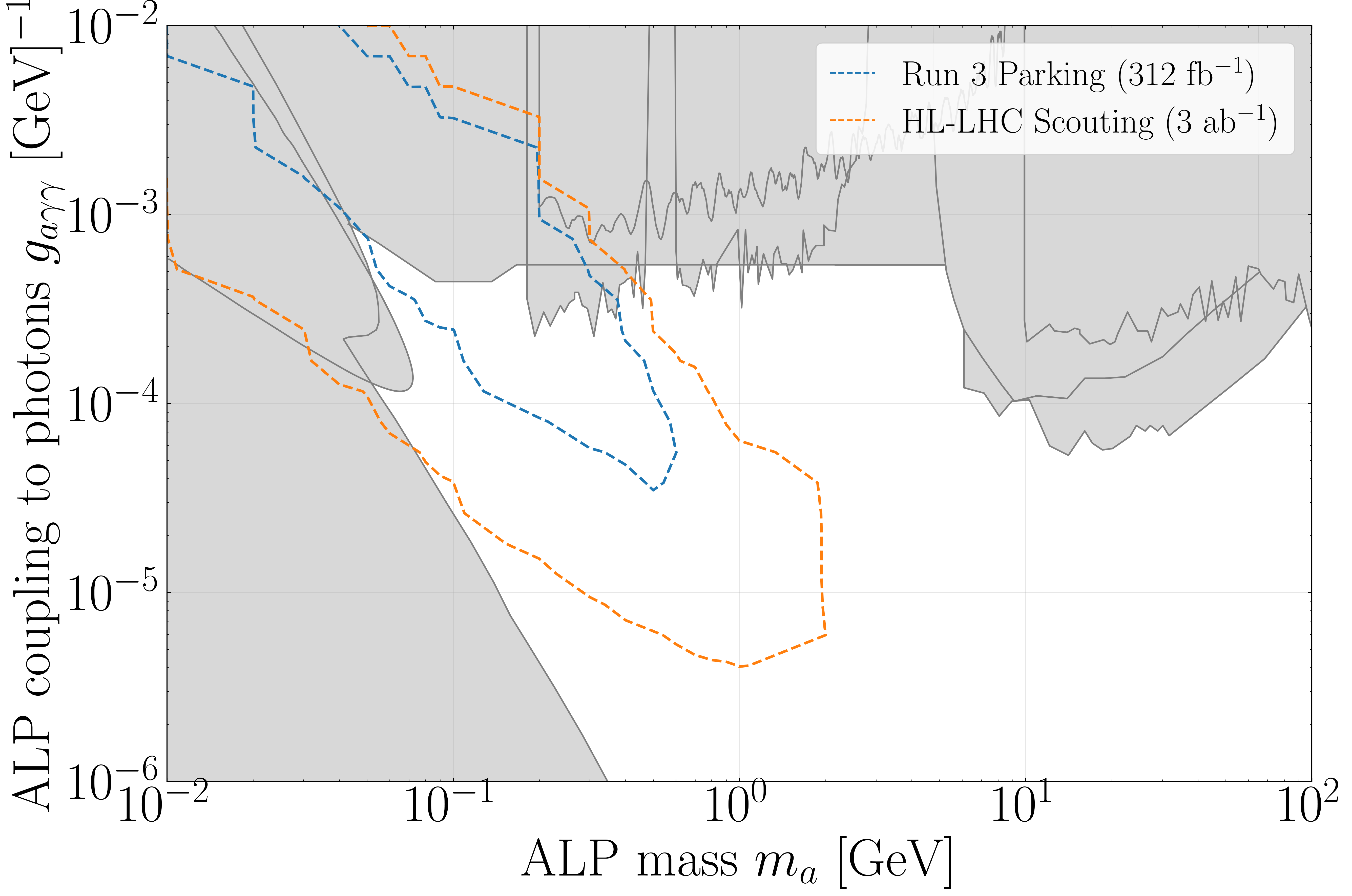}
\caption{Constraints on electromagnetically coupled ALPs. The colored contours
represent the regions accessible with the different approaches described in this
study. The gray area is excluded by other experiments (see text for details).
Contours are computed under a background-free assumption. This assumption does not
hold for $\ma \simeq m_{K_L} \simeq 0.5\GeV$ within the diphoton mass resolution,
where the residual $K_L\to\gamma\gamma$ background reduces the reach (Sec.~\ref{sec:bkg}).}
\label{fig:sensitivity}

\end{figure}
Combining the production cross section of Eq.~\eqref{eq:xsec}, the
pseudorapidity-dependent conversion probability of
Fig.~\ref{fig:conv-prob-combined}, and the acquisition strategies of
Secs.~\ref{sec:run3}--\ref{sec:phase2}, we project the CMS reach in the
$(\ma,\gagg)$ plane. Fig.~\ref{fig:sensitivity} shows the projected sensitivity for the configurations considered, spanning the Run~3
data-parking dataset ($312$\fbinv) and the HL-LHC L1T data scouting dataset
($3\,\mathrm{ab}^{-1}$).

Previous constraints in the parameter space $(m_a,\,g_{a\gamma\gamma})$ are shown in gray in Figs.~\ref{fig:parameter_space_variations} and \ref{fig:sensitivity}. 
In the low-mass regime, constraints arise from FASER~\cite{FASER:2024bbl} and a broad set of beam-dump experiments~\cite{Riordan:1987aw,CHARM:1985anb,Bjorken:1988as,Blumlein:1990ay,NA64:2020qwq}, with the strongest bounds currently provided by NA64~\cite{NA64:2020qwq}, NuCal~\cite{Blumlein:1990ay}, and E137~\cite{Bjorken:1988as}.
For intermediate masses, we show LEP exclusions~\cite{L3:1994shn,OPAL:2002vhf} as derived in Refs.~\cite{Jaeckel:2015jla,Knapen:2016moh}, along with Belle II~\cite{Dolan:2017osp,Belle-II:2020jti} and BESIII~\cite{BESIII:2022rzz,BESIII:2024hdv} limits. Earlier constraints from PrimEx~\cite{Aloni:2019ruo}, CLEO~\cite{CLEO:1994hzy}, BaBar~\cite{BaBar:2010eww}, and CDF~\cite{CDF:2013lma} have since been surpassed and are omitted for clarity.
For masses above a few GeV, the LHC provides the strongest experimental reach~\cite{dEnterria:2021ljz}. In particular, the large photon-photon luminosity available in ultra-peripheral Pb--Pb collisions~\cite{Knapen:2016moh,Knapen:2017ebd} allows CMS~\cite{CMS:2018erd} and ATLAS~\cite{ATLAS:2020hii} to set stringent limits. 
At still higher masses, a combination of dedicated LHC analyses~\cite{CMS:2011uvc,CMS:2011bsw,ATLAS:2011ab,CMS:2012cve,ATLAS:2012yve,ATLAS:2012fgo,ATLAS:2014jdv,ATLAS:2017ayi,CMS:2020rzi}, together with their reinterpretations~\cite{Jaeckel:2012yz,Jaeckel:2015jla,Knapen:2016moh,Knapen:2017ebd,Bauer:2017ris,Bauer:2018uxu,Florez:2021zoo}, extends the sensitivity to ALP masses well into the TeV range.

The HL-LHC scouting projection extends beyond the Run~3 parking projection over
essentially the whole mass range, reaching roughly an order of magnitude lower in
$\gagg$ and a comparable factor higher in $\ma$. Two effects drive this. First, the larger
dataset supplies the rate needed for the rarer, more weakly coupled signal. Second, the scouting stream imposes no trigger thresholds, so the soft signal characteristic of
small couplings is recorded in full rather than reduced by the VBF jet selection of
Table~\ref{tab:vbf_trigger}. The displaced-vertex requirement, common to both
strategies, rejects the prompt backgrounds. 
The Run~3 parking projection is therefore not the more sensitive of the two, but it is
the one that can be carried out on data already recorded. It sets the reach available
before the HL-LHC begins, and the scouting projection sets the reach available
afterward.

Both contours are computed under the assumption, stated in Sec.~\ref{sec:bkg}, that
the $K_L\to\gamma\gamma$ background is suppressed to a negligible level by the
strategies described there. In Run~3 this relies on the reconstruction of the merged
diphoton invariant mass. In the scouting stream, where no such reconstruction tool is yet available, it
relies on the diphoton $p_T$, the track-based quantities, and the hadronic
environment. The large $K_L$ sample that the scouting stream itself will provide makes it possible
to measure the residual background directly from data and to optimize these
discriminants against it. The HL-LHC contour therefore shows the sensitivity
that CMS can reach once such a $K_L$ rejection is in place for the scouting stream,
for example through a trigger-level mass estimator.

\section{Conclusions}
\label{sec:conclusions}
We have projected the sensitivity of the CMS experiment to light,
electromagnetically coupled ALPs in the $10\MeV$--$10\GeV$ mass range, produced
via VBF and decaying to collimated photon pairs reconstructed
through photon conversions in the silicon tracker. Adapting the
tracking-conversion methodology of~\cite{tracking-alps} to the CMS detector
and to the data-acquisition strategies available in CMS, we showed that the high trigger thresholds on photons can be
relieved by VBF-based data parking in Run~3 and by L1T data scouting at the
HL-LHC. Together, these strategies open access to a region of the $(\ma,\gagg)$ plane that
existing measurements have not reached. The breadth of acquisition strategies available to CMS
is what makes this reach possible. 

The projected reach rests on the assumption that the $K_L\to\gamma\gamma$ background,
which survives the displacement requirement, can be suppressed to a negligible level. This is the part of the program that
most needs further work. In Run~3, the decisive step is a reconstruction of the merged
diphoton invariant mass from the full ECAL and tracker information, for which existing
Transformer-based algorithms provide a starting point. The residual $K_L$ under the
signal peak can then be constrained from data. In the HL-LHC scouting stream, no such
tool is available yet, and the discrimination must be built from trigger-level quantities.
The diphoton $p_T$ provides strong separation on its own. The conversion tracks give
the individual photon momenta and their opening angle, from which a coarse mass
estimator could be constructed. The hadronic environment offers additional
discrimination, since the $K_L$ is produced inside a jet while the ALP recoils against
two forward quark jets and is otherwise isolated. Developing these into a working
selection, and quantifying the $K_L$ tail that our simulated samples cannot resolve,
are the natural next steps on the Phase-2 side, and the time remaining before HL-LHC data
taking starts leaves room to do so. 

Although VBF is the dominant production mode for electromagnetically coupled ALPs at the LHC, no CMS or ATLAS measurements have targeted it below 10 GeV. The parked Run~3 data on which such a search can be performed already exist. We expect the first LHC result on VBF-produced light ALPs to come from them.

\begin{acknowledgments}
GA acknowledges support from the Ram\'on y Cajal program, fellowship
reference RYC2024-048931-I funded by the MICIU/AEI/10.13039/501100011033
and the FSE+. JJ acknowledges support from the Deutsche Forschungsgemeinschaft (DFG, German Research Foundation)
under the Collaborative Research Centre SFB 1225 - 273811115 (ISOQUANT).
\end{acknowledgments}

\bibliography{APS_ref}
\end{document}